\documentclass[useAMS,usenatbib]{mnras}
\usepackage{graphicx}
\usepackage{longtable}
\usepackage{graphicx}
\usepackage{hyperref}
\usepackage{amsmath}

\title[Is the WELS class of PNe central star real?]
  {Problems for the WELS classification of planetary nebulae central stars: Self-consistent nebular
  modelling of four candidates.}
\author[H.M. Basurah et al.] {Hassan M. Basurah$^1$\thanks{Email: hbasurah@kau.edu.sa},
Alaa Ali$^{1,2}$,
Michael A. Dopita$^{1,3}$,
 \newauthor
R. Alsulami $^1$,
Morsi A. Amer$^{1,2}$,
 \& A. Alruhaili,$^1$ \\
  $^1$Astronomy Dept, Faculty of Science, King Abdulaziz University, Jeddah, Saudi Arabia \\
  $^2$Department of Astronomy, Faculty of Science, Cairo University, Egypt \\
  $^3$Research School of Astronomy and Astrophysics, Australian National University, Cotter Rd., Weston ACT 2611, Australia
  }
\date{Released XXXX Xxxxx XX}

\pagerange{\pageref{firstpage}--\pageref{lastpage}} \pubyear{2015}

\def\LaTeX{L\kern-.36em\raise.3ex\hbox{a}\kern-.15em
    T\kern-.1667em\lower.7ex\hbox{E}\kern-.125emX}

\begin{document}

\label{firstpage}

\maketitle

\begin{abstract}
 We present integral field unit (IFU) spectroscopy and self-consistent photoionisation modelling
 for a sample of four southern Galactic planetary nebulae (PNe) with supposed weak emission-line (WEL)
 central stars. The Wide Field Spectrograph (WiFeS) on the ANU 2.3 m telescope has been used to
 provide IFU spectroscopy for NGC 3211, NGC 5979, My 60, and M 4-2 covering the spectral range
 of 3400-7000\,{\AA}. All objects are high excitation non-Type I PNe, with
strong He II emission, strong [Ne\,V]
 emission, and weak low-excitation lines. They all appear to be predominantly optically-thin
 nebulae excited by central stars with $T_{\rm eff} > 10^5$\,K. Three PNe of the sample have
 central stars which have been previously classified as weak emission-line stars (WELS), and the
 fourth also shows the characteristic recombination lines of a WELS. However, the
 spatially-resolved spectroscopy shows that rather than arising in the central
 star, the C\,IV and N\,III recombination line emission is distributed in the
 nebula, and in some cases concentrated in discrete nebular knots. This may suggest that
 the WELS classification is spurious, and that, rather, these lines arise from (possibly
 chemically enriched) pockets of nebular gas. Indeed, from careful background subtraction
 we were able to identify three of the sample as being hydrogen rich O(H)-Type. We have
 constructed fully self-consistent photoionization models for each object. This
allows us to independently determine the chemical abundances in the
nebulae, to provide new model-dependent distance estimates, and to
place the central stars on the H-R diagram. All four PNe have
similar initial mass ($1.5 < M/M_{\odot} <2.0$) and are at a similar
evolutionary stage.
\end{abstract}


\section{Introduction}

Planetary nebulae represent an advanced stage in the evolution of
low and intermediate mass stars as they make the transition between
the asymptotic giant branch (AGB) and the white dwarf (WD) stages.
The gaseous nebula which appears now as a planetary nebula (PN) is
the remnant of the deep convective envelope that surrounded the
central core of AGB. This core is now revealed as the central star
(CS) of the PN. Thus the PNe provide fundamental data on the
mass-loss processes during the AGB stage, the chemical enrichment of
the envelope by dredge-up processes, as well as information about
the mass, and effective temperature of the remaining stellar core.

Up to now, most studies of the emission line spectra of PNe have
been derived from long-slit spectroscopic work. However, an accurate
derivation of the physical conditions and chemical abundances in the
nebular shell relies upon a knowledge of the integrated spectra. For
Galactic PNe, this can only be determined using integral field
spectroscopic instruments. This field was pioneered by
\citet{Monreal-Ibero05} and \citet{Tsamis07}, although it is only
recently that detailed physical studies using optical integral field
data have been undertaken, amongst which we can cite
\citet{Monteiro13} (NGC 3242), \citet{Danehkar13} (SuWt 2),
\citet{Danehkar14} (Abell 48), \citet{Danehkar15} (Hen 3-1333 and
Hen 2-113) and \citet{Ali15b} (PN G342.0-01.7). In this paper, we
will analyse integral field data obtained with the  WiFeS integral
field spectrograph \citep{Dopita07,Dopita10} to derive plasma
diagnostics, chemical composition, and kinematical parameters of the
four highly excited nebulae NGC 3211, NGC 5979, My 60, and M 4-2 and
to study the properties of the central stars (CSs) of these objects,
which from the evidence of their spectra alone, appear to belong to
the class of weak emission-line stars (WELS), \citet{Tylenda93},
whose properties are described below.

Reviewing the literature, none of our target objects have been
subject to an individual detailed study. However, many of their
nebular properties and central star characteristics have been
derived and the results are distributed amongst a large number of
articles, particularly in the cases of NGC 3211 and NGC 5979. From
imaging data, the nebula NGC 3211 was classified as an elliptical
PNe \citep{Gorny99} with an extended halo of radius $\sim$ 3.2 times
that of the main nebular shell \citep{Baessgen89}. The plasma
diagnostics and elemental abundances of the object were studied in a
few papers \citep{Perinotto91,Milingo02}. The central star of NGC
3211 has several temperature and luminosity estimates: 135\,kK \&
2500\,L$_{\odot}$  \citep{Shaw89}; 155\,kK \& 345\,L$_{\odot}$
\citep{Gathier89} and 150\,kK \& 1900\,L$_{\odot}$
\citep{Gruenwald95}. \citet{Gurzadian88} provided upper and lower
estimates of CS temperature of 89\,kK and 197\,kK, respectively.

The morphology of NGC 5979 was studied clearly from the narrow band
[O III] image taken by \citet{Corradi03}. The object was classified
as elliptical PN surrounded by a slightly asymmetrical halo. Another
[O III] image from the HST was taken by \citet{Hajian07} shows a
similar structure, but at better resolution. \citet{Phillips00}
suggested that such halos are likely arise through the retreat of
ionization fronts within the nebular shell, as the CS temperature
and luminosity decline at intermediate phases of PN evolution. The
CS of NGC 5979 was classified as a WELS by \citet{Weidmann11}, but
later \citet{Gorny14} claimed that the key CS emission lines
features of the WELS type;  C IV at 5801-12 \AA{} and N III, C III,
and C IV complex feature at 4650 \AA{} can be of nebular origin.
\citet{Stanghellini93} reported remarkably small values of both the
temperature (58\,kK) and luminosity (71\,L$_{\odot}$) of NGC 5979
central star. However these estimates were based on the HI Zanstra
temperature method, which can be considerably in error if the nebula
is optically thin. More likely values for the CS temperature
(100\,kK) and luminosity (14000\,L$_{\odot}$) were given by
\citet{Corradi03}. The details of the chemical composition of NGC
5979 have been studied in the optical regime by \citet{Kingsburgh94}
and \citet{Gorny14}.

\citet{Ruffle04} presented an $H\alpha$ contour map of My 60 that
shows a circular or only mildly elliptical morphology of the object.
From the velocity measurements of the nebular shells,
\citet{Corradi07} found that one side of the nebular shell has lower
expansion velocity than the opposite side. \citet{Stanghellini93}
reported that My 60 is an elliptical PN with multiple shell and its
central star has an effective temperature of 113 kK and luminosity
of 4216 L$_{\odot}$. A detailed spectroscopic study of My 60 and its
CS was recently made by \citet{Gorny14}, where they derived the
nebular chemical composition and classified its CS as a WELS type.

\citet{Weidmann11} presented a low resolution spectrum for the CS of
the M 4-2 nebula which revealed the characteristic emission lines of
WELS type. For this object, \citet{Zhang93} reported a stellar
temperature of 101kk and a luminosity of 5600 L$_{\odot}$. The
H$\alpha$ and [O III] images presented by \citet{Schwarz92}, reveal
a roughly circular nebula with unresolved internal structure. Weak
constraints on  the nebular temperature, density, and chemical
composition were given by \citet{Kaler96}.

 Our interest in studying this particular group of PNe is that all four appear to be members of the (carbon rich)
 WELS class of central stars. The WELS denomination was proposed by spectral characteristics of this class was
 described in detail by \citet{Marcolino03}, who argue that there seems to be a general evolutionary sequence
 connecting the H-deficient central stars and the PG\,1159 stars which link the PNe stars to white dwarfs.
 The proposed evolutionary sequence originally developed by \citet{Parthasarathy98} is [WCL] $\rightarrow$
 [WCE] $\rightarrow$ WELS $\rightarrow$  PG\,1159.

\citet{Fogel03} were unsuccessful in finding an evolutionary
sequence for WELS similar to what had been established for the [WR]
CS. However, they found that WELS have intermediate stellar
temperature (30-80 kK). They found no WELS associated with Type I
PNe -
 all studied objects having N/O ratios lower than 0.8,
indicating lower mass precursors. However, we should note here that the lower limit
of the N/O ratio in Type I PNe is 0.5 as defined by \citet{Peimbert83}
and \citet{Peimbert87}. Adopting this criteria of Type I PNe to the
analysis of \citet{Fogel03}, would imply that $\sim 20\%$ (5 objects) of the
sample (26 objects) are of Type I (Figure 2, \citet{Fogel03}).
\citet{Girard07} affirm, on average, WELS have slightly lower helium
and nitrogen abundances compared to [WR] and non-WR PNe. They find
somewhat enhanced helium and nitrogen abundances in [WR] PNe with an
 N/O ratio $\sim 4$ times solar value, while WELS have N/O
ratios which are nearly solar value. From the IRAS two-colour
diagram, they find that WELS are shifted to bluer colours than the
other [WR] PNe. \citet{Frew12} show that WELS have larger scale
height compared to other many CS classes and consequently they rise
from low mass progenitor stars.

 The emission lines which characterise the WELS spectral
 class are the recombination lines of C and N and consist of the following: N\,III $\lambda \lambda 4634,
 4641$,  C\,III $\lambda 4650$, C\,IV $\lambda 4658$, and C\,IV $\lambda \lambda  5801, 5812$. These lines
 are indistinguishable in width from the nebular lines, although on low dispersion spectra the group of
 lines around 4650\AA\ and the C\,IV doublet around 5805\AA\ may appear to be broad features.
 \citet{Frew12} noted that the characteristics WELS
recombination lines such as C\,IIII, C\,IIV and N\,IIII were also
observed in some massive O-type stars, as well as in low-mass X-ray
binaries and cataclysmic variables. \citet{Corradi11} observed these
C\,IIII, C\,IIV  and N\,IIII emission lines in the spectrum of the
close binary CS of the high-excited PN IPHASX J194359.5+170901.
Also, \citet{Miszalski11} explained that many of the characteristic
WELS emission lines have been observed in close binary central stars
of PNe such as the high-latitude planetary nebula ETHOS 1. Further,
they note that lines originate from the irradiated zone on the side
of the companion facing the primary. On this basis
\citet{Miszalski09} claim that many of WELS are likely to be
misclassified close binaries.

It has been difficult to characterise the exact evolutionary state
or nature of the WELS class.
 For example, from UV data, \citet{Marcolino07} find lower terminal velocities than those characterising
 the [WC] - PG\,1159 stars, arguing that the latter form a distinct class. \citet{Weidmann15} used
 medium-resolution spectra taken with the Gemini Multi-Object Spectrograph (GMOS) to discover
 that 26\% of of them are H-rich O stars, and at least 3\% are H deficient. They argue against the denomination
 of WELS as a spectral type, since the low-resolution spectra generally used to provide this classification do
 not provide enough information about the photospheric H abundance. In addition, we have the disturbing conclusion
 by \citet{Gorny14} that in NGC 5979 the strong recombination line emission arises not in the central star, but in the surrounding nebula.

The literature on the WELS class will be more extensively discussed
below, but it is clear from the above discussion that high spectral
resolution integral field data should cast light on the evolutionary
status of the WELS. Such data are presented in this paper. The
observations and the data reduction are described in Section
\ref{Obs}, while the physical conditions, ionic and elemental
abundances determinations are given in Section \ref{Parameters}.
Section \ref{Velocities} is dedicated to study the expansion and
radial velocities of the sample. In Section \ref{CS}, we discuss the
evidences that prove the PNe central stars can not be classified as
WELS type. A full discussion on the global photoionisation models is
provided in Section \ref{Models}.

\section{The Integral Field Observations}\label{Obs}
\subsection{Observations}
The integral field spectra of the PNe were obtained over two nights
of March 30-31, 2013 with the Wide Field Spectrograph;
\citep{Dopita07, Dopita10} mounted on the 2.3-m ANU telescope at
Siding Spring Observatory \citep{Mathewson13}. The WiFeS instrument
delivers a field of view of 25\arcsec x 38\arcsec at a spatial
resolution of either 1.0\arcsec x 0.5\arcsec or 1.0\arcsec x
1.0\arcsec, depending on the binning on the CCD. In these
observations we used the 1.0\arcsec x 1.0\arcsec option. The blue
spectral range of 3400-5700 {\AA}  was covered at a spectral
resolution of $ \sim 3000$ using the B3000 grating, that corresponds
to a full width at half maximum (FWHM) of $\sim 100$ km/s. In the
red, the R7000 grating was used, which covers the spectral range of
5700-7000 {\AA} at a higher spectral resolution $R \sim 7000$
corresponding to a FWHM of $\sim 45$ km/s. The wavelength scale was
calibrated using the Cu-Ar arc Lamp with 40s exposure at the
beginning and throughout the night,  while flux calibration was
performed using the STIS spectrophotometric standard stars HD 111980
\& HD 031128 \footnote{Available at : \newline {\tt
www.mso.anu.edu.au/~bessell/FTP/Bohlin2013/GO12813.html}}. In
addition a B-type telluric standard HIP 38858 was observed to
correct for the OH and H$_2$O telluric absorption features in the
red. The separation of these features by molecular species allowed
for a more accurate telluric correction which accounted for night to
night variations in the column density of these two species.

\subsection{Data Reduction}
All the data cubes were reduced using the PyWiFeS \footnote {http://www.mso.anu.edu.au/pywifes/doku.php.} data reduction
pipeline \citep{Childress14}. A summary of the on-target spectroscopic observation log is shown in Table \ref{Table1}.
In the long exposures,  some strong nebular emission lines such as [O III] $\lambda$5007 and H$\alpha$ are saturated on
the CCD chip. For these, the fluxes were measured from the additional short exposure time observations.

The resultant data cubes are summed from the original exposures, cleaned of cosmic ray events, wavelength calibrated to
better than 0.4\AA\ in the blue, and 0.07\AA\ in the red, corrected for instrumental sensitivity in both the spectral and
spatial directions, sky background subtracted, and telluric absorption features have been fully removed. On the basis of
the intrinsic scatter in derived sensitivity amongst the various standard star observations we can estimate the error in
absolute flux of these observations to be $\pm 0.015$\,dex throughout the wavelength range covered by these observations.

\begin{table*}
\centering
 \small
   \caption{A Summary of the observing log}
   \label{Table1}
   \scalebox{0.95}{
  \begin{tabular}{llllllll}
 \hline
   Nebula name  & PNG number & no. of  & Exposure  & Date   & Airmass  & $v_{hel}$ corr. & Standard Stars Used \\
    & & frames & time (s) &  &  &km/s & \\
   \hline \hline
 NGC 3211 & PN G286.3-04.8    &  6 & 100 & 31/3/2014 &1.19 & 2.98 &  HD 111980 \& HD 031128 \\
 &    & 3 & 600 & 31/3/2014 & 1.2 & 2.98 & HD 111980 \& HD 031128 \\
 NGC 5979 & PN G322.5-05.2    &  6 & 300 & 30/3/2014&1.25 & 20.0 & HD 111980 \& HD 160617 \& HD 031128 \\
 &    & 3 & 600 & 30/3/2014 & 1.18& 20.0 &HD 111980 \& HD 160617 \& HD 031128 \\
 My 60 & PN G283.8+02.2       &  3 & 200 & 30/3/2014 &1.12 & 1.5 &HD 111980 \& HD 160617 \& HD 031128 \\
 &    & 3 & 600 & 30/3/2014 &1.1 & 1.5 & HD 111980 \& HD 160617 \& HD 031128 \\
 M 4-2 & PN G248.8-08.5      &  6 & 300 & 30/3/2014 & 1.0 & -14.8 & HD 111980 \& HD 160617 \& HD 031128 \\
 \hline
 \end{tabular}}
\end{table*}

\subsection{Images}\label{Images}
We extracted continuum-subtracted emission line images from the data cubes using {\tt QFitsView v3.1 rev.741}
\footnote{{\tt QFitsView v3.1} is a FITS file viewer using the QT widget library and was developed at the Max
Planck Institute for Extraterrestrial Physics by Thomas Ott.}.
These are then used to construct three-colour images in any desired combination of emission lines.

\begin{figure*}
\centerline{\includegraphics[scale=0.6]{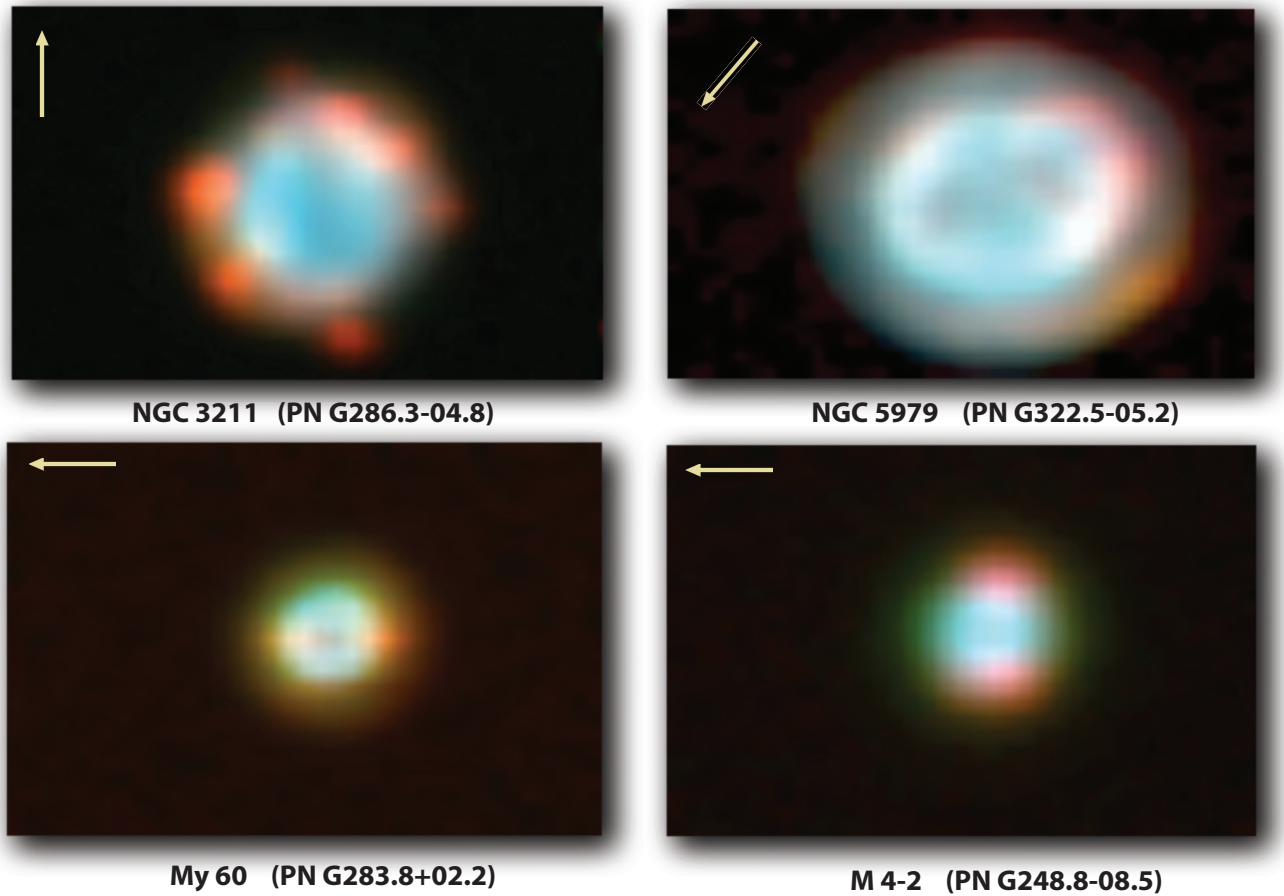}}
\caption{Narrow band continuum-subtracted image composites of our
PNe derived from the WiFeS data cubes. The field size of each image
is 25\arcsec x 38\arcsec. The yellow arrow denotes the direction of
north in each image. The colour coding of these images is He II
$\lambda4686$ (blue), H$\beta$ (green) and [N II] $\lambda6583$
(red), and they are presented on a square root stretch to bring out
the fainter features. Each image has been boxcar smoothed by 1
spaxel or 1.0\arcsec. to remove sharp spaxel boundaries. Note that
all objects have only fragmentary shells of [N II] emission, which
indicates that all four objects are optically thin to the escape of
ionising radiation.}\label{fig1}
\end{figure*}

To illustrate the overall morphology and excitation structure of
these PNe, we present in Figure \ref{fig1} images in HeII (blue
channel), H$\beta$ (green channel) and [N II] (red channel). These
three ions probe the excitation structure and the overall
distribution of gas in these nebulae. It is clear from the
incomplete shell morphology in the [N II] $\lambda6563$ line that
all of these PNe are optically-thin to the escape of EUV photons. In
the cases of NGC\,3211 and NGC\,5979, the outer shell is clumpy and
broken, while in the other two cases the low-excitation gas is
distributed symmetrically on either side of the nebula. These cases
are reminiscent of the fast, low-ionization emission regions in
NGC\,3242, NGC\,7662, and IC\,2149 and other PNe described in a
series of papers \citep{Balick93, Balick94, Hajian97, Balick98}, and
discussed on a theoretical basis by \citet{Dopita97}. In the case of
My 60, the existence of FLIERs may well be the cause of the
asymmetric expansion reported by \citet{Corradi07}. The double shell
structure of NGC\,5979 is clearly evident.

\subsection{Global Spectra}\label{Spectra}
In order to build a detailed photoionisation model of the PN, a mean
spectrum of the whole nebula is required. From the 3-D data cube, we
extracted this spectrum using a circular aperture which matched the
observed extent of the bright region of the PNe using {\tt
QFitsView}. The residuals of the night sky lines of [O\,I]
$\lambda$5577.3, [O\,I] $\lambda$6300.3 and [O\,I]$ \lambda$6363.8
were removed by hand. The scaling between the blue and red spectra
caused by slight mis-match of the extraction apertures was improved
by measuring the total fluxes in the common spectral range between
blue and red channels (5500-5700 {\AA}), and applying the scale
factor to the red spectrum to match the two measured fluxes.

We measured emission-line fluxes from the final combined,
flux-calibrated blue and red spectra of each PN. The line fluxes and
their uncertainties were measured using the IRAF {\tt
splot}\footnote{IRAF is distributed by the National Optical
Astronomy Observatory, which is operated by the Association of
Universities for Research in Astronomy (AURA) under a cooperative
agreement with the National Science Foundation.} task and were
integrated between two given limits, over a local continuum fitted
by eye, using multiple Gaussian fitting for the line profile.  We
used the Nebular Empirical Abundance Tool (NEAT\footnote{The code,
documentation, and atomic data are freely available at
http://www.sc.eso.org/~rwesson/codes/neat/.}; Wesson et al. 2012) to
derive the reddening coefficients  and in the subsequent plasma
diagnostics and ionic and elemental abundances calculations. The
derived logarithmic reddening coefficients and the H$\beta$ fluxes
are given in Table \ref{Table2}, and the complete list of line
intensities is given in the appendix in Table \ref{TableA1}.

\begin{table*}
\centering \caption{Reddening coefficients, observed $H\beta$
fluxes, excitation classes and inferred distances of our sample.}
\label{Table2} \scalebox{1.0}{
\begin{tabular}{lclclccccc}
 \hline
      {\bf Object}        & \multicolumn{2}{|c|}{$ {\bf c(H\beta)}$} & \multicolumn{2}{|c|}{\bf Log F$(H\beta)$} &
      \multicolumn{2}{|c|}{\bf EC} &  \multicolumn{2}{|c|}{\bf Distance (kpc)}\\
               & This Paper & Literature & This paper & Literature  & (6) & (7) & (8) & (9) \\
 \hline
NGC 3211       & 0.37     & 0.34(1), 0.32(2), 0.25(3)  & -11.12  & -11.06(2), -11.39(3)  &  10.1 & 9.1   & 2.61 & 2.59\\
NGC 5979       & 0.40     &  0.38(1),  0.25(4) & -11.34  & -11.23(2), -11.9(5) & 10.8 & 10.1   & 2.99 & 2.49\\
My 60          & 0.86     & 0.91(1), 0.95(2), 0.87(4)  & -11.83  & -11.78(2), -11.8(5) & 10.4 & 8.8   & 3.76 & 3.33\\
M 4-2          & 0.57     & 0.37(1)  & -11.85  & -11.70(5) & 9.5 &9.0 &  5.91 & 5.72  \\
 \hline
\end{tabular}}
\begin{minipage}[!t]{16cm}
{\small {\bf References.}: (1) \citet{Tylenda92};(2) \citet{Shaw89};
(3) \citet{Milingo02}; (4) \citet{Gorny14}; (5) \citet{Acker91}; (6)
\citet{Reid10}; (7) \citet{Meatheringham91};
(8) \citet{Ali15a}; (9)  \citet{Frew16}, assuming all objects are optically thin.\\
}
\end{minipage}
\end{table*}

\section{Determination of physical conditions}\label{Parameters}

\subsection{Temperatures and densities}
The spectral lines in the sample given in Table \ref{TableA1} allow
us to determine the electron densities from the low and medium
ionization zones and the electron temperature from the low, medium
and high ionization zones. The  lines of low ionisation were used to
determine the nebular densities from the [S~II]
$\lambda$6716/$\lambda$6731 and [O~II] $\lambda$3727/$\lambda$3729
and the temperatures from [N~II] ($\lambda 6548 + \lambda
6584$)/$\lambda$5754 line ratios, while in the more highly ionised
zones we determine the nebular densities from the [Cl~III]
$\lambda$5517/$\lambda$5537,  [Ar~IV] $\lambda$4711/$\lambda$4740
and [Ne~IV] $\lambda$4715/$\lambda$4726 line ratios, while  the
temperature is estimated from the  [O~III] ($\lambda 4959 + \lambda
5007$)/$\lambda$4363 line ratio. For NGC 5979, we also able to
determine the nebular temperature from the high ionization line
ratio [Ar~V] ($\lambda 6435 + \lambda 7005$)/$\lambda$4625.  The
Monte Carlo technique was used by NEAT to propagate the statistical
uncertainties from the line flux measurements through to the derived
abundances. In Table \ref{Table3}, we list the inferred nebular
densities, temperatures and their uncertainties for each object. In
addition, we provide comparisons of our results with those which
have previously appeared in the literature. In general, we find good
agreement with other results.

Table \ref{Table3} also gives the values of the density and
temperature derived from the global photoionisation  models which
are presented in detail and discussed in Section \ref{Models},
below.

\begin{table*}
\centering
\caption{Electron temperatures and densities measured for the PNe sample, and fitted using global photoionisation modelling.}
\label{Table3}
\scalebox{0.85}{
\begin{tabular}{|p{2.5cm}|p{1.5cm}|p{1.5cm}|p{1.0cm}|p{2.0cm}|p{1.5cm}|p{1.5cm}|p{1.5cm}|p{1.5cm}|p{1.0cm}|}
 \hline
      {\bf Object }        & \multicolumn{4}{|c|}{\bf Temperature (K)} & \multicolumn{5}{|c|}{\bf Density (cm$^{-3}$)} \\
               & {[}O III{]} & {[}N II{]} &  {[}S II{]} & {[}Ar V{]} & {[}S II{]} & {[}O II{]} & {[}Ar IV{]} & {[}Cl III{]} & {[}Ne IV{]} \\
                \hline
{\bf NGC 3211}  &   &    &    &     &    &  &   &  \\
Measured:      & 13937$\pm$ 162     & 11892$\pm$587  &   &  & 1363$\pm$270     & 1601$\pm$134       & 1583.6$\pm$317      & 1048$\pm$848  &  \\
Nebular Model:         & 14810     & 11910  &   10910 &  17040     & 1370       & 1684      & 1464  & 1538 & 1310 \\
Ref (1) & 13300     & 10300  &    &      & 1200       &       &   &  \\
Ref (2) & 14320     &   &    &      &        &       &   &  \\
Ref (3) & 14000     & 12500  &    &      &        &  970     &   &  \\
\hline
{\bf NGC 5979}  &   &    &    &     &    &  &   &  \\
Measured:      & 13885$\pm$234     & 13578$\pm$1217    & &14375$\pm$902  & 1470$\pm$320     & 2034$\pm$261     & 1342$\pm$503      & 1342$\pm$503  &  \\
Nebular Model:         &   13860   &  13640 &  13690  &   15180    &  1680      &   1695    &  1625 &  1667 & 1583\\
Ref (4) & & & &  & 1549 & 1585 & 1862 & 2570 \\
Ref (5) &      & $14604^{+15206}_{-14049}$  &    &      & $509^{+849}_{-240}$       &       &   &  \\
Ref (6) & 13100 & & & & & 1410 & & &  \\
\hline
{\bf My 60}  &   &    &    &     &    &  &   &  \\
Measured:      & 13665$\pm$229     & 13411$\pm$688   & & & 2098$\pm$215     & 3009$\pm$706     & 1763$\pm$472      & 1637$\pm$945 &      \\
Nebular Model:   &13210      &  13270    & 13340  &  15250  &   2132   &  2150     & 2045  & 2113 & \\
Ref (5) &       & $13868^{+14412}_{-13324}$  &    &      & $1510^{+2450}_{-1050}$       &       &   &  \\
\hline
{\bf M 4-2}  &   &    &    &     &    &  &   &  \\
Measured:          & 15170$\pm$186     & 11084$\pm$380   & & & 1662$\pm$78     & 2178$\pm$235     & 1537$\pm$486        & 1164$\pm$864  &     \\
Nebular Model:         &  14070    & 12080  & 11350   &   16870    &  1775      &  2088     & 1680  & 2027 &  1745\\
 \hline
\end{tabular}}
\begin{minipage}[!t]{16cm}
{\small {\bf References:} (1) \citet{Henry04}; (2) \citet{Kaler96};
(3) \citet{Liu93};   (4) \citet{Wang04}; (5) \citet{Gorny14}; (6)
\citet{Kingsburgh94}}
\end{minipage}
\end{table*}

\subsection{Excitation classes and distances}
The detected emission lines in all objects cover a wide range of
ionization states from neutral species such as [O I], all the way up
to [Ne V] which requires the presence of photons with greater energy
than 97.1\,eV. For medium and high excitation classes (EC), above
EC=5, the line ratio of He II $\lambda 4686${\AA}$/H\beta$ provides
the best indication of excitation class of the nebula. This line is
only present when the central star has an effective temperature,
$T_{\rm eff} > 85000$K. Based on the analysis of 586 PNe in the
Large Magellanic Cloud (LMC), \citet{Reid10} compared the three main
classification schemes of \citet{Aller56}, \citet{Meatheringham91}
and \citet{Stanghellini02} to estimate the excitation classes of
PNe. They noticed that at EC $\geq 5$, the line ratio
[O~III]$/H\beta$ is not fixed but it increases to some degree with
nebular excitation. Therefore they introduced a new method to
incorporate both the He II $\lambda 4686${\AA}$/H\beta$ and
[O~III]$/H\beta$ line ratios in classifying the EC of the nebula.
Here we present (Table \ref{Table2}) the excitation class for the
four objects using both methods of \citet{Reid10} and
\citet{Meatheringham91}.

Determining a reliable distance for a planetary nebula is not an
easy task. To determine the distance of PN, one should rely first on
the individual methods, then on the statistical methods. A summary
on the applications, assumptions, limitations, and uncertainties  of
individual methods was given in \citet{Ali15a} and \citet{Frew16}.
Considering the individual distances in the literature, we found
that NGC 3211 and NGC 5979 have each two distance estimates, and My
60 has only one. NGC 3211 has distances of 1.91 kpc
(\citet{Gathier86} - extinction method) and 3.7 kpc
(\citet{Zhang93b} - gravity method). The latter technique should be
the most reliable in this case.  NGC 5979 has distances of 2.0 kpc
(\citet{Frew16} - expansion method) and 5.8 kpc (\citet{Zhang93b} -
gravity method). My 60 has a distance of 3.2 kpc (\citet{Zhang93b} -
gravity method). On average, \citet{Ali15a} finds the gravity method
overestimates the PN distance by $\sim 25\%$ compared to other
methods. It is also obvious here that a large variation exists
between the various distance estimates of NGC 3211 and NGC 5979.
This discrepancy in distance estimates commonly appears, not only
among different methods, but sometimes between different authors
using the same method \citep{Ali15a}. Lacking trusted individual
distances, e.g. trigonometric or cluster membership distances, for
any of the sample, we rely initially on the statistical methods. For
each PN in our sample, we derived two statistical distances which
are given in Table \ref{Table2}. The first value was deduced from
the distance scale of \citet{Ali15a}, while the second value was
taken from that of \citet{Frew16}. We find small distance variations
between both distance scales ($\sim 1\% - 17\%$), assuming that all
objects are optically thin PNe (Table \ref{Table2}). In this paper
we have adopted the nebular distances derived from the distance
scale of \citet{Ali15a}. These distances may be compared with those
derived in Section 6.2 (below).

On the basis of these distances, we derived absolute
luminosities $L_{\rm H\beta}$: $1.44\times10^{34}$~erg~s$^{-1}$,
$1.24\times10^{34}$~erg~s$^{-1}$, $1.79\times10^{34}$~erg~s$^{-1}$
and $2.20\times10^{34}$~erg~s$^{-1}$ for NGC 3211, NGC 5979, My 60,
and M 4-2 respectively.

\subsection{Ionic and elemental abundances}
Applying the NEAT, ionic abundances of nitrogen, oxygen, neon,
argon, chlorine and sulfur were calculated from collisional
excitation lines (CEL), while helium and carbon were calculated from
optical recombination lines (ORL) using the temperature and density
appropriate to their ionization potential. When several lines from a
given ion are present, the ionic abundance adopted is found by
averaging the abundances from each ion and weighted according to the
observed intensity of the line. The total abundances were calculated
from ionic abundances using the ionization correction factors (ICF)
given by \citet{Kingsburgh94}, to correct for unseen ions.

The total helium abundances for all objects were determined from
He$^+$/H and He$^{2+}$/H ions.  The total carbon abundances were
determined from C$^{2+}$/H and C$^{3+}$/H ions for all objects, except
My 60 which is determined from C$^{2+}$/H ion only, using ICF (C) =
1.0.

None of the objects studied here has He/H $\geq 0.125$ and/or N/O
$\geq 0.5$.
Therefore, they can not be classified as a Type I according to the
original classification scheme proposed by \citet{Peimbert78}. This
probably implies that their progenitor stars had lower initial
masses ($M < 4M_{\odot}$). According to the derived chemical
compositions of the objects, they will most likely to be classified
as Types II and III. These types of PNe have abundances which more
nearly reflect the properties of the interstellar medium out of
which their central stars have been formed, particularly with
respect to those chemical elements that are not contaminated by the
evolution of intermediate mass stars, such as oxygen, neon, sulphur
and argon. The He/H, N/H and N/O abundances in NGC 3211 and NGC 5979
are consistent with Type IIa PNe \citep{Quireza07}. The Galactic
vertical height (z $\leqslant$ 1 kpc) and peculiar velocity
($\leqslant$ 60 km$s^{-1}$) of NGC 3211 confirm that it is probably
a member of Galactic thin disk. The chemistry of My 60 and M 4-2 is
consistent with Type IIb/III PNe, both nebulae have Galactic
vertical heights less than 1 kpc, but peculiar velocities larger
than 60 km s$^{-1}$.  Therefore we suggest that both objects are of
Type III PNe, which are usually located in the Galactic thick disk.
These results are consistent with that of \citet{Fogel03} and
\citet{Girard07} where none of WELS are associated with Type I PNe.

Tables \ref{Table4} and \ref{Table5}, compare the abundance
determinations of the four objects with those obtained by our
detailed global photoionisation models and with those given by other
authors.

\begin{table*}
\centering
\caption{Abundances derived from the NEAT and MAPPINGS codes for NGC 3211 and NGC 5979. Model I
refers to the total abundances in the photoioionsation model, while model II refers to gas phase abundances in the same model.}
\label{Table4}
\scalebox{0.85}{
\begin{tabular}{lccccccccccc}
 \hline
  & \multicolumn{6}{|c|}{\bf NGC 3211} & &\multicolumn{4}{|c|}{\bf NGC 5979} \\ \cline{2-7} \cline{9-12}
Element & Neat & Model I & Model II & Ref 1 & Ref 2 & Ref 3 & & Neat & Model I & Model II & Ref 4\\
\hline
He/H & 1.06E-1$\pm$2.7E-3 & 1.10E-1 & 1.10E-1
& 1.10E-1 & & 1.17E-1&
& 1.10E-1$\pm$3.9E-3 & 1.10E-1 & 1.10E-1
& 1.11E-1$^{+1.17{\rm E}-1}_{-1.05{\rm E}-1}$ \\
C/H & 1.20E-3$\pm$6.7E-5 & 1.34E-3 & 8.64E-4
&  &  & &
& 2.05E-3$\pm$1.3E-4 & 1.84E-3 & 1.19E-3
&  \\
N/H & 1.15E-4$\pm$9.9E-6 &  7.96E-5 &  6.62E-5
& 1.63E-4 &   & &
& 1.38E-4$\pm$1.2E-5 & 2.81E-5  &  2.33E-5
& 6.70E-5$^{+7.62{\rm E}-5}_{-5.62{\rm E}-5}$ \\
O/H & 4.43E-4$\pm$2.0E-5 & 4.60E-4 & 2.97E-4
& 8.38E-4 & 5.00E-4& 8.77E-4 &
& 4.55E-4$\pm$3.0E-5 & 6.21E-4& 3.74E-4
& 3.44E-4$^{+4.06{\rm E}-4}_{-2.87{\rm E}-4}$ \\
Ne/H & 9.30E-5$\pm$3.9E-6 & 7.82E-5 & 7.82E-5
& 1.31E-4 & 1.38E-4 & &
& 5.63E-5$\pm$7.8E-6 & 6.05E-5 & 6.05E-5
& 5.75E-5$^{+6.59{\rm E}-5}_{-5.13{\rm E}-5}$   \\
Ar/H & 1.90E-6$\pm$9.6E-8 & 1.60E-6 & 1.60E-6
 & 6.33E-6 & & &
 & 2.17E-6$\pm$1.3E-7 &  2.17E-6& 2.17E-6
 & 2.27E-6$^{+2.48{\rm E}-6}_{-2.03{\rm E}-6}$ \\
K/H &  & 8.07E-7 & 8.07E-7  &
&  &   & & \\
S/H & 6.18E-6$\pm$4.83E-7 & 5.41E-6 & 5.41E-6
& 4.40E-6 & 1.34E-5 & & & 8.48E-6$\pm$7.3E-7 &  3.92E-6& 3.92E-6
& 5.32E-6$^{+5.92{\rm E}-6}_{-4.62{\rm E}-6}$  \\
Cl/H & 1.46E-7$\pm$1.44E-8 & 5.40E-7 & 1.45E-7
& 4.52E-7 & & &
& 1.90E-7$\pm$1.6E-8 & 4.20E-7& 1.13E-7
& 9.30E-6$^{+1.13{\rm E-5}}_{-6.87{\rm E-6}}$ \\
Fe/H &   &  & 6.80E-8 &
&   & & & & & 2.97E-8\\
N/O & 0.26  &  0.17 &    0.23 &  & &    &
& 0.23  &   0.06 &     0.08  &   \\
\hline
\end{tabular}}
\begin{minipage}[!t]{14cm}
Note: The abundances of K and C elements in models I \& II were
calculated from [K IV] and C II $\lambda$4267 emission lines,
respectively. References: (1) \citet{Henry04}; (2) \citet{Maciel99};
(3) \citet{Liu93}; (4) \citet{Gorny14}.
\end{minipage}
\end{table*}

\begin{table*}
\centering \caption{Abundances derived from the NEAT and MAPPINGS
codes for My 60 and M 4-2. Model I refers to the total abundances,
while model II refers to gas phase abundances in the same model.}
\label{Table5} \scalebox{0.9}{
\begin{tabular}{lccccccccc}
 \hline
  &   \multicolumn{4}{|c|}{\bf My 60}  &   & \multicolumn{4}{|c|}{\bf M 4-2}\\ \cline{2-5} \cline{7-10}
Element & Neat & Model I & Model II & Ref 1 &  & Neat & Model I & Model II & Ref 2 \\
\hline
He/H
& 1.10E-1$\pm$3.9E-3 & 1.10E-1 & 1.10E-1 & 1.10E-1$^{+1.16{\rm E}-1}_{-1.03{\rm E}-1}$ &
& 9.81E-2$\pm$2.7E-3 & 1.04E-1 & 1.04E-1 & 1.14E-1\\
C/H
& 6.31E-4$\pm$1.2E-4 & 1.16E-3 & 7.50E-4 & &
& 2.58E-3$\pm$1.2E-4 & 1.71E-3 & 1.19E-3 \\
N/H &
6.85E-5$\pm$7.5E-6 &  4.98E-5 &  4.14E-5 & 5.11E-5$^{+6.40{\rm E}-5}_{-4.17{\rm E}-5}$ &
& 5.73E-5$\pm$3.8E-6 & 5.74E-5  & 5.24E-5 & 1.97E-4\\
O/H
& 3.15E-4$\pm$2.0E-5 & 4.60E-4 & 2.97E-4 & 3.15E-4$^{+3.80{\rm E}-5}_{-2.62{\rm E}-4}$ &
& 2.40E-4$\pm$1.1E-5 & 2.08E-4 & 1.62E-4& 4.70E-4\\
Ne/H
& 6.20E-5$\pm$3.9E-6 & 6.81E-5 & 6.81E-5 & 6.02E-5$^{+6.92{\rm E}-5}_{-5.24{\rm E}-5}$ &
& 4.85E-5$\pm$2.1E-6 &  4.26E-5& 4.26E-5\\
Ar/H
 & 1.59E-6$\pm$1.1E-7 & 2.20E-6 & 2.20E-6 & 1.68E-6$^{+1.86{\rm E}-6}_{-1.48{\rm E}-6}$ &
 & 9.27E-7$\pm$1.5E-8 & 8.30E-7 & 8.30E-7 \\
K/H
&  & & 1.01E-7 &
&  &  & 1.90E-7\\
S/H &
4.92E-6$\pm$4.7E-7 & 4.50E-6 & 4.50E-6 & 4.33E-6$^{+5.10{\rm E}-6}_{-3.66{\rm E}-6}$ &
& 2.35E-6$\pm$1.5E-7 & 3.80E-6 & \\
Cl/H &
1.20E-7$\pm$1.4E-8 & 4.01E-7 & 1.10E-7 & 8.33E-6$^{+1.04{\rm E}-5}_{-6.34{\rm E}-6}$ &
& 4.48E-8$\pm$4.1E-9 &  9.31E-8 & \\
Fe/H
&   &  & 9.62E-8 & &
&   &  & 6.75E-8\\
N/O
& 0.22  &   0.11       &     0.14    &     &
& 0.24  &   0.22       &    0.26          \\
 \hline
\end{tabular}}
\begin{minipage}[!t]{14cm}
Note: The abundances of K and C elements in models I \& II were
calculated from the [K\,IV] and C\,II $\lambda$4267 emission lines,
respectively. References: (1) \citet{Gorny14}; (2) \citet{Kaler96}.
\end{minipage}
\end{table*}

\section{Expansion and radial velocities}\label{Velocities}

The expansion velocity is an essential quantity to determine in
order to understand nebular evolution. We used three emission lines
([S\,II], [N\,II], [Ar\,V]) that all lie in the red part of the
WiFeS spectrum, and are consequently observed at higher spectral
resolution (R=7000) than the blue spectra (R=3000). These lines were
used to determine the nebular expansion velocity $V_{\rm exp}$ for
each object. The full width at half maximum (FWHM) of each line was
measured using the IRAF {\tt splot} task. The full width was
corrected for instrumental and thermal broadening to derive the
expansion velocity using the following formula \citep{Gieseking86};
\begin{equation}\label{3}
  V_{\rm exp} = 0.5 \left[w^2_{\rm obs} - w^2_{\rm inst} - 8(\ln 2)kT_e/m\right]^{1/2}
\end{equation}
where $w_{\rm obs}$ is the observed FWHM of the measured line,
$w_{\rm inst}$ is the instrumental FWHM, $k$ Boltzmann's constant,
$T_e$ is the nebular temperature, and $m$ the atomic mass of the
species emitting the measured line. The results are given in Table
\ref{Table6}. As expected from the ionisation stratification of
these PNe, the measured expansion velocity decreases with increasing
ionisation potential of the ion. Apparently, there are no previous
attempts to measure the expansion velocity of M 4-2. NGC 3211 has
measured velocities of 26.5 kms$^{-1}$, 31.0 kms$^{-1}$ and 21.1
kms$^{-1}$ using the [O\,III], [O\,II] and He II emission lines,
respectively \citep{Meatheringham88}. Our derived expansion
velocities for NGC 3211 in the [N\,II] and [S\,II] are smaller than
those of \citet{Meatheringham88}, except for He II line. This may be
attributed to the derivation of the expansion velocity from an
integrated spectrum over the whole object, while other authors
usually derive the expansion velocity at a certain nebular position
using long slit spectra. In the study of \citet{Meatheringham88},
they measured the maximum velocity for asymmetric objects by set the
spectrograph slit along the nebular long axis. It is clear also from
the NGC 3211 colour image in Figure \ref{fig1} that the [N\,II]
emission is strongly concentrated in blobs located in the outer
regions of the nebula. \citet{Hajian07} gave an equatorial expansion
velocity of 18 kms$^{-1}$ for NGC 5979 derived from long slit
spectra in the [O\,III] emission line. This value is consistent with
our measurements and slightly smaller than that derived from [S\,II]
and [N\,II] emission lines, due to the higher ionisation potential
of the [O\,III] line. \citet{Schonberner05} provided two expansion
velocities for the rim of My 60.  They derived $V_{\rm exp}$ = 23.1
kms$^{-1}$ and 23.7 kms$^{-1}$ from [O\,III] and [N\,II] emission
lines, respectively. The latter one is in good agreement with our
derived value.

The systemic velocities $RV_{\rm sys}$ of the sample were measured
using the IRAF external package {\tt RVSAO}. A weighted mean radial
velocity was calculated for each object using $H\alpha$, [N~II], and
[S~II] emission lines. The heliocenteric correction was taken from
the image header, to derive the heliocentric radial velocity
$RV_{\rm hel}$ for each object.  The results were listed in Table
\ref{Table6}. The value of NGC 3211 is in good agreement with
\citet{Durand98} and \citet{Schneider83}. The radial velocity of NGC
5979 is higher than that derived by \citet{Schneider83}, $RV_{\rm
hel}= 23\pm3.0$, which was taken originally from the low dispersion
spectra of \citet{Campbell18}. Therefore our larger value of
$RV_{\rm hel}$ for NGC 5979, compared to that of
\citet{Schneider83}, may be attributed to the different spectral
resolution as well as to the fact that we derived the radial
velocity from the integrated spectrum over the whole object, while
other value was derived from a particular nebular position.

\begin{table*}
\centering \caption{Radial and expansion velocities of the sample.}
\label{Table6}
\begin{tabular}{|p{1.5cm}||p{1.5cm}|p{4cm}|p{0.9cm}||p{0.9cm}|p{0.9cm}|}
 \hline
  {\bf Object}    & \multicolumn{2}{c}{\bf $RV_{\rm hel}$ (km/s)}&
 \multicolumn{3}{|c|}{\bf V$_{exp}$ (km/s)} \\  \cline{4-6}
      &    This paper    &   Literature     & [S II] & [N II] & [Ar V] \\
\hline
NGC 3211       & -21.8$\pm$3.5 & -22.3$\pm$1.6(1), -22.5$\pm$2.8(2) & 23.6     & 22.0     & 10.3   \\

NGC 5979       & 30.4$\pm$3.5  & 23.0$\pm$3.0(1)    & 20.7      & 22.4     & 11.8 \\
My 60          & 42.2$\pm$3.0 &    & 24.2     & 24.1     & $<$ 10 \\
M 4-2          & 119.1$\pm$4.7   &  & 25.9      & 25.5    & 9.3 \\
 \hline
\end{tabular}
\begin{minipage}[!t]{13cm}
{\tiny {\bf References.} : (1) \citet{Durand98}; (2) \citet{Schneider83} \\
}
\end{minipage}
\end{table*}

\section{What are the central stars?}\label{CS}

\begin{figure*}
\includegraphics[scale=0.65]{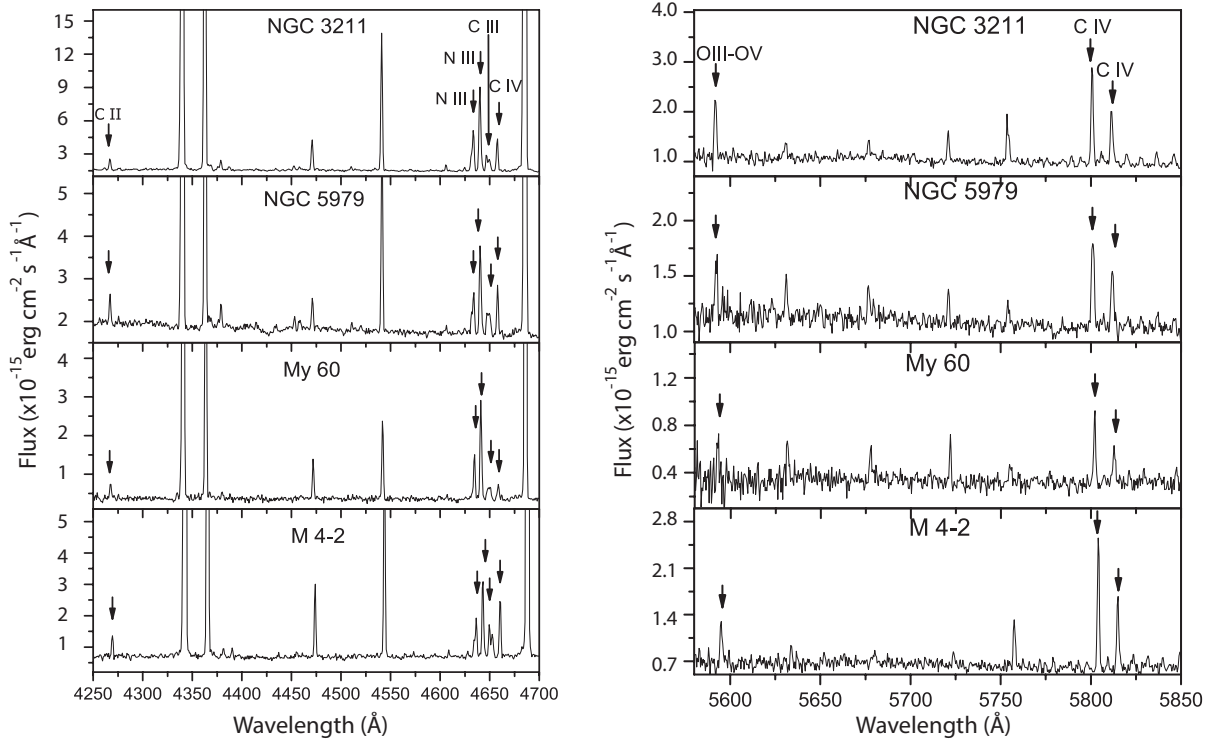}
 \caption{Characteristic emission lines generally used to classify central stars as of the WELS type, as seen in our PNe sample. }
  \label{fig:Fig2}
\end{figure*}

\begin{figure*}
 \caption{Narrow band continuum-subtracted image composites of our PNe in the recombination lines of C\,IV $\lambda5801$
 (blue), N\,III $\lambda 4641$  (green), [N II] $\lambda6583$ (red). As in Figure \ref{fig1}, they are presented on
 a square root stretch to bring out the fainter features and boxcar smoothed by 1 spaxel or 1.0 arc sec. to remove
 sharp spaxel boundaries. The arrow shows the direction of north. No trace of a central star is seen in any object.
 Instead, the recombination line emission of C\,IV  and N\,III is - in each case - entirely of nebular origin.}
  \label{fig3}
\end{figure*}

Of the over 3000 Galactic PNe currently known \citep{Frew10}, only
13 \%  of their stellar nuclei have been spectroscopically
identified \citep{Weidmann11}. According to the observed atmospheric
spectra of the CSs of PNe, \citet{Mendez91} has assigned two types.
In the first type hydrogen is the prevalent element (H-rich), while
the second type is relatively free of hydrogen and the spectra are
dominated by lines of  helium and carbon elements (H-deficient).

In general, the H-deficient type can be divided into three main
groups. The first group, the [WR] group, shows spectra with strong
and broad emission lines mainly from He, C, and O, similar to the
Wolf-Rayet stars of population I. This group was divided into three
subgroups: [WCL], [WCE], and [WO], according to the ionization
stages of the element dominating the atmosphere \citep{Crowther98,
Acker03}. The second group, PG1159, is composed of pre-white dwarfs
that have mainly absorption lines of He II and C IV in their spectra
(Werner et al. 1997). The third group, [WC]-PG 1159, includes
objects with similar optical spectra to the  PG1159 stars, but have
strong P Cygni lines in the UV (e.g. N\,V $\lambda$1238 and C\,IV
$\lambda$1549).

\citet{Tylenda93} presented the analyses of extensive observations
of a set of 77 emission-line CSs. About half of this set were
classified as [WR] stars and the other half exhibited emission lines
at the same wavelengths as [WR] stars, but having weaker intensities
and narrower line widths. These they named ``weak emission-line
stars'' (WELS). The spectra of the WELS show the carbon doublet,
C\,IV $\lambda$5801 and $\lambda$5812 and the $\lambda$4650 feature
which is a blend of  N\,III $\lambda$4634 and $\lambda$4641, C\,III
$\lambda$4650, and C\,IV $\lambda$4658. Furthermore the spectra of
WELS are highly ionized to the degree that the C\,IV
$\lambda$$\lambda$5801, 5812 doublet is strong but the C\,III
$\lambda$5696 line is either very weak or absent. Based on the
analysis of the optical spectra of 31 WELS, \citet{Parthasarathy98}
found that the spectrum of a WELS is very similar to those of the
[WC]-PG1159 and PG1159 stars. As pointed out in the introduction,
they suggested that an evolutionary sequence exists connecting the
[WC] central stars of PNe to the PG1159 pre-white dwarfs: [WC late]
$\rightarrow$ [WC early] $\rightarrow$ WELS = [WC]-PG1159
$\rightarrow$ PG1159. Based on the UV spectra of the WELS,
\citet{Marcolino07} disagreed with \citet{Parthasarathy98}, and
instead claimed that the WELS are distinct from [WC]-PG1159 stars.
They noticed that WELS type presented P Cygni features and most of
their terminal velocities lie in the range from $\sim$1000-1500 km
s$^{-1}$, while [WC]-PG1159 stars have much higher values of
$\sim$3000 km s$^{-1}$. In addition, they found that the [WC]-PG1159
stars are characterised by several intense P Cygni emission in the
$\sim$1150-2000\AA{} wavelength interval, most notably N\,V
$\lambda$1238, O\,V $\lambda$1371, and C\,IV $\lambda$1549, while in
the WELS O\,V $\lambda$1371 is either very weak or absent.
Furthermore, \citet{Parthasarathy98} and \citet{Marcolino03} find
some absorption lines in the optical spectra of WELS such as He\,II
$\lambda$4541 and $\lambda$5412 indicating that any stellar wind is
not as dense as in the case of [WR] stars.

\citet{Hajduk10} claimed that not all WELS spectral type are
H-deficient as [WR] stars, but some may be H-rich despite having
emission lines. Recently, \citet{Weidmann15} have observed and
studied 19 WELS from the total of 72 objects known in literature.
The high quality spectra of these objects show a variety of spectral
types, where 12 of them have H-rich atmospheres, with different wind
densities and only 2 objects seem to be H-deficient. They were not
able to decide on the spectral type of the remaining objects, but
they concluded that these objects are not [WR] stars. According to
the above results, they claimed that  ``the denomination WELS should
not be taken as a spectral type, because, as a WELS is based on
low-resolution spectra, it cannot provide enough information about
the photospheric H abundance''.

Three objects of our sample have previously been associated with the
WELS type: NGC 5979, M 4-2 \citep{Weidmann11} and My 60
\citep{Gorny14}. Due to the low spectral resolution, the components
of C+N $\lambda$4650 feature were not resolved in the spectra
presented by \citet{Weidmann11} - their Figure 11 and
\citet{Gorny14} - their Figure C.2. Furthermore the components of
the doublet C\,IV centred at 5805\AA{} were not resolved in the
spectra given by \citet{Weidmann11} - their figure 11.  In Figure
\ref{fig:Fig2}, we illustrate our observations of the characteristic
WELS emission lines seen in all four of our objects. Here, to
simulate the effect of observing with a long-slit spectrograph, we
extracted these spectra by taking a circle of 2\arcsec radius around
the geometric centre of each object.  In general, the spectra of all
nebulae roughly show similar emission lines, generally identified to
be of the WELS class;  C\,II at 4267\AA{}, N\,III at 4634\AA{} and
4641\AA{} , C\,III at 4650\AA{} (here resolved into its two
components at 4647\AA{} and 4650\AA{}),
 C\,IV at 4658\AA{}, O\,III-O\,V at 5592\AA{}, and finally, C\,IV at
5801\AA{} and 5812\AA{}.

\citet{Gorny14} noticed that it was possible that - as in the case
of NGC\,5979 - the CS spectrum can ``mimic'' the WELS type. They
found that the key CS emission features of WELS type (C II at
4267\AA{}, N III at 4634\AA{} and 4641\AA{}, C III at 4650\AA{}, C
IV at 5801\AA{} and 5812\AA{}) appear in a spatially extended region
in the 2D spectra of NGC 5979, and therefore, they are of nebular
rather than stellar origin.

To test whether the CSs in our objects were real WELS-type, or
simply ``mimics'', we constructed continuum-subtracted images in the
brightest recombination lines  C\,IV $\lambda5801$  and  N\,III
$\lambda 4641$, and formed colour composites using the procedure
described in Section \ref{Images}. The result is shown in Figure
\ref{fig3}. From this is evident that the recombination lines are
formed in the nebula, and that the spatial separation of the C\,IV,
N\,III and [N\,II] emission is entirely consistent with ionisation
stratification. We therefore conclude that, in every case the
emission is entirely of nebular origin, and that the WELS
classification is spurious for these objects.

\begin{figure*}
\includegraphics[scale=0.6]{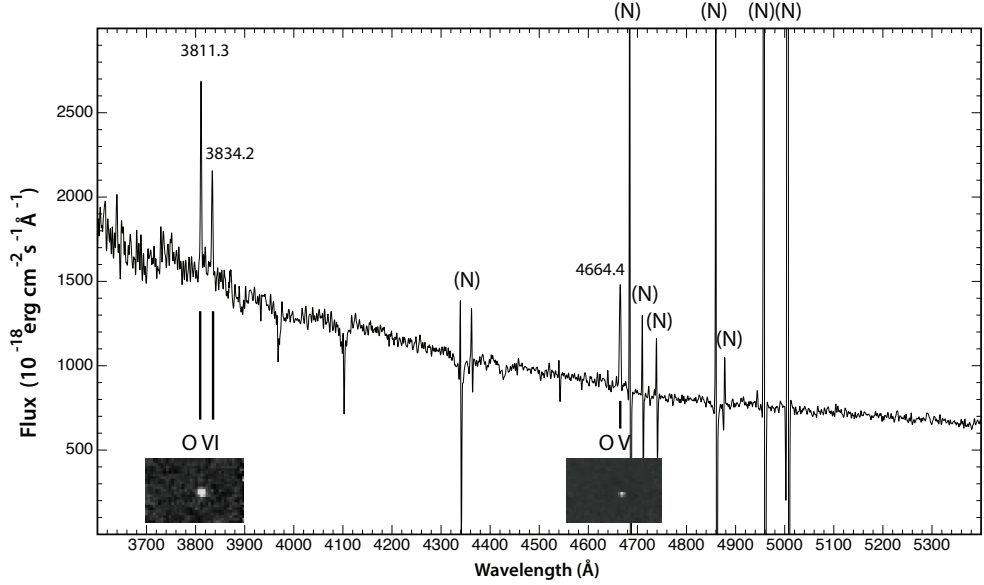}
 \caption{The nebular subtracted spectrum of the central star in NGC\,5979 (showing the full WiFeS image field). Emission
 lines of O\,VI and O\,V are clearly detected in the central star, as indicated in the
 monochromatic thumbnail images shown in the lower part of the figure. The features marked (N)
 are improperly subtracted nebular features. There are broad Balmer absorption features present
 in the blue continuum of the star.}
  \label{fig4}
\end{figure*}

Certainly, the CSs are detected, but if not of the WELS type, what
could they be? In order to examine this, we need to carefully remove
the nebular emission. This was possible in three cases; NGC\,3211,
NGC\,5979 and My 60. The result is shown in Figure \ref{fig4} for
NGC\,5979. There are broad Balmer absorption features present in the
blue continuum of the star, and three stellar emission lines, O\,VI
$\lambda\lambda 3811.3, 3834.2$ and O\,V $\lambda 4664.4$
{\footnote{The identification of this line was determined from the
Kentucky database {\tt http://WWW.pa.uky.edu/$\sim$peter/atomic/}}.
The presence of broad Balmer absorption in the CS proves the
presence of H in the atmosphere, and establishes that the CS is
likely to be H-burning. However, the source of the narrow O\,VI and
O\,V lines remains uncertain. Being narrow, they most likely have
their origin in or close to the photosphere, rather than in the
wind. We tentatively classify this star, following the CS
classification scheme of \citet{Mendez91}, as O(H)-Type.} These
features clearly originate in the central star as can be seen from
the continuum-subtracted emission line maps shown as thumbnail
images in Figure \ref{fig4}. In the case of NGC\,3211 and My 60,
only the O\,VI $\lambda\lambda 3811.3, 3834.2$ doublet was
identified in their CSs, but this is sufficient to classify these
stars as also belonging to the O(H)-Type. In M 4-2 only a faint blue
continuum could be identified as coming from the central star, so
this star remains without classification.

\section{Self-consistent photoionization modelling}\label{Models}
Integral field spectroscopy of PNe in our Galaxy presents an ideal
opportunity to build fully self-consistent photoionisation models.
Not only can a global spectrum be extracted which is directly
comparable to the theoretical model, but we are able to use the size
and morphology to constrain the photoionisation structure and inner
and outer boundaries of the nebula. In addition, we can directly
compare measured electron temperatures and densities in the
different ionisation zones to constrain the pressure distribution
within the ionised material. In this section, we describe how such
self-consistent photoionisation modelling can provide a good
description to all these parameters, while at the same time enabling
us to derive chemical abundances, to derive distances, and to place
the central stars on the H-R diagram. We refer the reader to earlier
work of this nature on the PN NGC\,6828 by \citet{Surendiranath08}.

We have used the {\tt Mappings 5.0} code (Sutherland et al. 2015, in
prep.) {\footnote{Available at {\tt miocene.anu.edu.au/Mappings}}}.
Earlier versions of this code have been used to construct
photoionisation models of H\,II regions, PNe, Herbig-Haro Objects,
supernova remnants and narrow-line regions excited by AGN. This code
is the latest version of the {\tt Mappings 4.0} code earlier
described in \citep{Dopita13}, and includes many upgrades to both
the input atomic physics and the methods of solution.

We choose for the initial abundance set  in the models a value of
0.8 times local galactic concordance (LGC) abundances based upon the
\citet{Nieva12} data on early B-star data. The  \citet{Nieva12} data
provide the abundances of the main coolants, H, He, C, N. O, Ne, Mg,
Si and Fe and the ratios of N/O and C/O as a function of abundance.
For the light elements we use the \citet{Lodders09} abundance, while
for all other elements the abundances are based upon
\citet{Scott15a, Scott15b} and \citet{Grevesse10}. The individual
elemental abundances are then iterated from this initial set.

The elemental depletions onto dust grains must be taken into
account. Not only do these remove coolants from the nebular gas, but
in PNe they are also an important source of photoelectric heating
\citep{Dopita00}. For the depletion factors we use the
\citet{Jenkins09} scheme, with a base depletion of Fe of  1.5\,dex.
For the dust model, we use a standard \citet{MRN} dust grain size
distribution.

For the spectral energy distribution of the CS, we use the
\citet{Rauch03} model grid, which provide metal-line blanketed NLTE
model atmospheres for the full range of parameters appropriate to
the CSs of PNe. Our initial estimate uses the $\log g = 6.0$ models.

The pressure in the ionised gas is an important parameter. In the
modelling we have used the isobaric approximation (which includes
any effect of radiation pressure), and we have matched the pressure
to fit the electron densities derived from observation, given in
Table \ref{Table3}, above. In this table we have also listed the
electron densities (and electron temperatures) returned from our
``best fit'' models. Generally speaking, the agreement is good to a
few percent.

It is clear from the optical morphology given in Figure \ref{fig1}
that all four of our nebulae are optically thin to the escape of
ionising photons, some more than others. We therefore developed a
2-component model consisting of the weighted mean of an
optically-thin component and an optically-thick component.

In addition to the chemical abundance set, the gas pressure, and the
fraction of optically-thin gas $F$, the parameters which determine
the relative line intensities in the model are the stellar effective
temperature $T_{\rm eff}$, the ionisation parameter at the inner
boundary of the PNe, $U_{\rm in}$ and the fraction of the
Str\"omgren radius ($R_{\rm S}$) at which the outer boundary of the
optically-thin component is located $f_{\rm S}=R_{\rm out}/R_{\rm
S}$. The objective of the modelling is to produce a unique solution
to each of these variables.

The strength of the emission lines of low excitation such as
[N\,II], [S\,II] and [O\,II] is very sensitive to the fraction of
optically-thick gas present, and these therefore provide an
excellent constraint on $F$. A further constrain on  $F$ is
furnished by the [O\,III]/H$\beta$ ratio, since this first
increases, and then sharply decreases as the  nebula becomes
increasingly optically-thin. All four of our PNe are located close
to the maximum in the [O\,III]/H$\beta$ ratio that is observed in
PNe. In addition, the helium and hydrogen excitation as measured by
the He\,II$\lambda4686$/He\,l$\lambda5876$ and the
He\,l$\lambda5876$/H$\beta$ ratios is sensitive to both $T_{\rm
eff}$ and $U_{\rm in}$. For a given $f_{\rm S}$, there is a narrow
strip of allowed solutions with $\log U_{\rm in}$ falling as $T_{\rm
eff}$  rises.

 In order to measure the goodness of fit of any particular photoionisation model,
 we measure the L1-norm for the fit. That is to say that we measure the modulus of the
 mean logarithmic difference in flux (relative to H$\beta$) between the model and the observations \emph{viz.};
\begin{equation}
{\rm L1} ={ \Sigma}_n \left | \log \left[ {F_n({\rm model})} \over
{F_n({\rm obs.)}} \right]\right | /n, \label{L1}
\end{equation}
where n is the number of lines being considered in the fit. This
weights fainter lines equally with stronger lines, and is therefore
more sensitive to the values of the input parameters. (By
themselves, the stronger lines would not provide sufficient
constraints on the variables of the photoionisation models, and the
signal to noise in our spectra is very adequate to measure these
fainter lines with sufficient accuracy - see Table A1). Typically,
we simultaneously fit between 27 and 35 emission lines.

For a given  $f_{\rm S}$ we ran an extensive grid of models in $\log
U_{\rm in}$ and  $T_{\rm eff}$, and searched for the region where
the L1-norm was minimised at the same time as the helium excitation
matching the observed value. When this point was identified, we
iterated the abundance set in a restricted $\log U_{\rm in}$ and
$T_{\rm eff}$ range to improve on the L1-norm. Typically, we could
reduce this to about 0.05\,dex, the error being mostly dominated by
a few lines for which apparently the models are inadequate. These
are discussed below.
\begin{figure}
\includegraphics[scale=0.5]{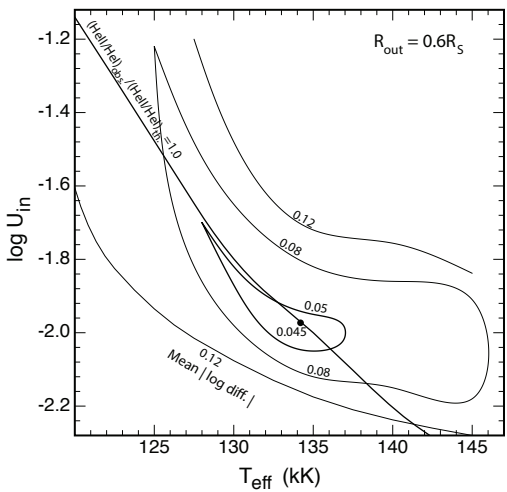}
 \caption{The best fit solution for M 4-2 is shown here on the $\log U_{\rm in}$ and  $T_{\rm eff}$ plane.
 We seek to minimise the L1-norm (contours), here given by fitting 35 lines, while simultaneously
 matching the He\,II, He\,I and H\,II excitation to that observed in the nebula.}
  \label{fig5}
\end{figure}

\begin{figure*}
\includegraphics[scale=0.9]{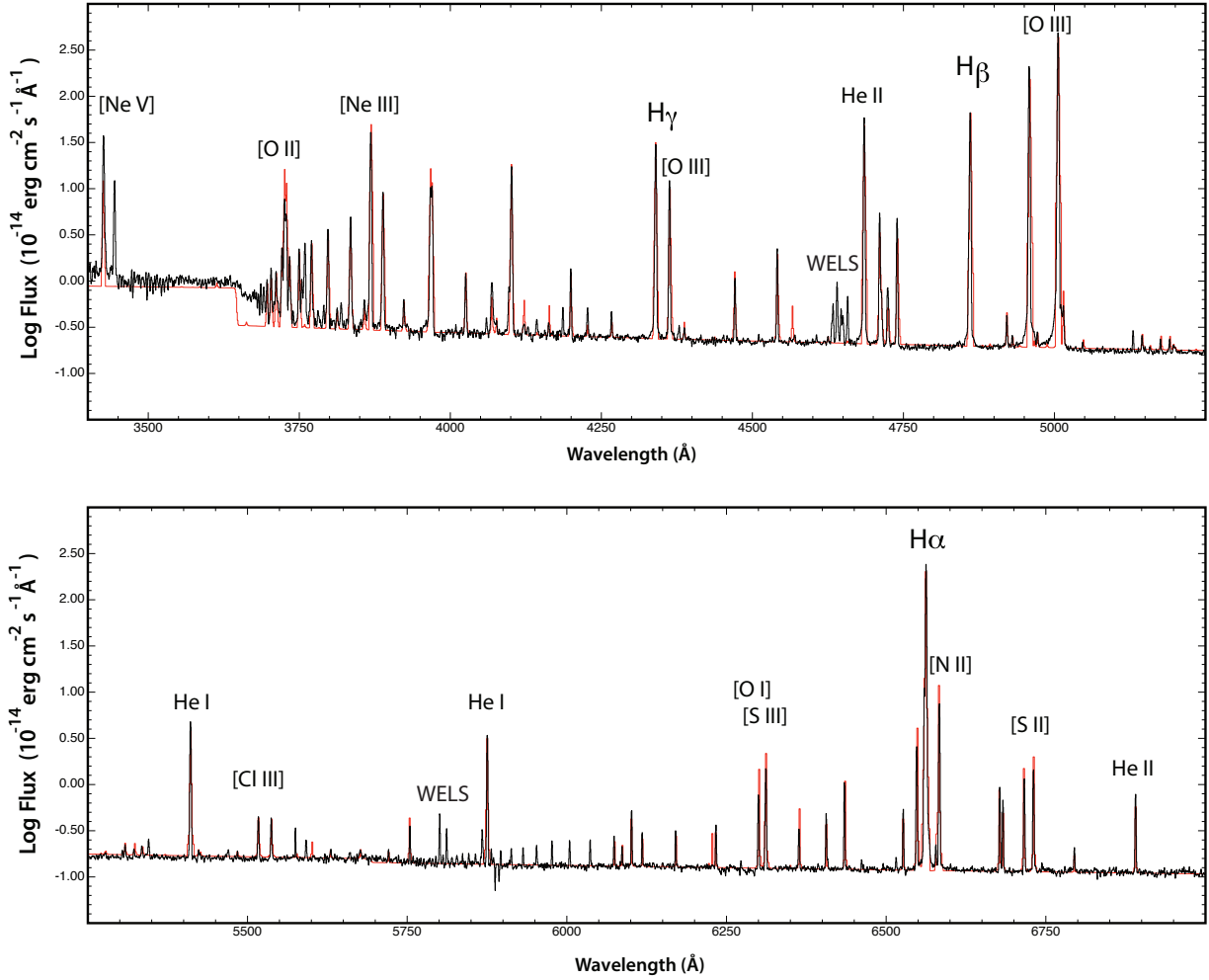}
 \caption{The comparison between the observed (dereddened) global spectrum of M 4-2
 (PN G248.8-08.5) (black) and the best fit  {\tt Mappings 5.0} photoionisation model (red).
 Some of the principal emission lines are identified. Note the excellent agreement between
 the observed and the theoretical nebular + stellar continuum, which is only possible
 when the observed spectrum has been correctly de-reddened.}
  \label{fig6}
\end{figure*}

In Figure \ref{fig5} we show the solution in the case of M 4-2. For
this model, we also provide in Figure \ref{fig6} a comparison
between the theoretical model (including the central star) and the
observations. Here we have matched the resolution of the theoretical
spectrum to that of the observations, as best as we can. Note that
the flux scale is logarithmic and covers nearly 4\,dex. The  {\tt
Mappings 5.0} code does not provide predictions for the higher
Balmer and He\,II $n=5$ series, as these become sensitive to the
density. Nor does the code predict the intensities of the
recombination lines of highly ionised species such as the N\,III,
C\,III, C\,IV, O\,V and O\,VI lines discussed above. This is simply
because the appropriate recombination coefficients and other atomic
data are not available for these transitions in these ions. Normally
the recombination coefficients into the excited states would be
estimated from the photoionization cross-sections using the Milne
relation. However, there are no published estimates of
photoionization cross-sections from these excited states.
Furthermore, the population of the upper level is also determined by
the cascade from more highly excited states, as well as by
collisional contributions from the lower levels. Finally, for some
excited levels radiative transfer effects (e.g. case A or B) can
make large differences to the predicted recombination line
strengths, and UV pumping into upper states followed by radiative
cascade into the upper state of the transition being considered can
also be very important. Apart from these issues, which regrettably
specifically affect the characteristic WELS recombination lines, the
overall fit of the model to the observation is excellent.

The derived parameters for the``best fit''  final models are as
follows:

\begin{tabular}{|l|c|c|c|c|c|}
  \hline
   & log P/k  & $F$  & $f_{\rm S}$ & log $U_{\rm in}$ & $T_{\rm eff}$ \\
      & cm$^{-3}$K &   &  &  & kK\\
\hline \\
 NGC 3211 & 7.63 & 0.992 & 0.72 & -1.95 & 145 \\
 NGC 5979 & 7.62 & 1.00 & 0.64 & -1.60 & 160 \\
  My 60 & 7.72 & 1.00 & 0.60 & -1.55 & 130 \\
  M 4-2 & 7.73 & 0.935 & 0.60 &-1.90 & 134 \\
  \hline
\end{tabular}

In Table \ref{Table7}, we present a comparison of the observed and
modelled line intensities of the lines used in the fitting. Overall
the agreement between the model and the observations is good.
However, we note a systematic effect that limits the minimum value
of the L1-norm. Although the predicted [Ne\,III] line intensities
are satisfactory, the [Ne\,V] lines are too weak with respect to the
[Ne\,IV] lines, while at the same time the [Ar\,V] lines are too
strong with respect to the [Ar\,IV] lines. The ionisation potential
of Ne\,III is 63.45eV that of Ne\,IV is 97.12eV, and the ionisation
potential of Ar\,IV is 59.81eV. These are all high, but in the same
general range. Thus, the differences between the predictions and the
observations cannot probably be ascribed to differences in the input
ionising spectra, but are more likely caused by errors in the charge
exchange reactions rates used by the code, which will affect the
detailed ionisation balance in the nebula.

\begin{table*}
\centering \caption{Line fluxes, relative to H$\beta = 100$, from
photoionisation models compared with those observed in NGC 3211, NGC
5979, My 60, and M 4-2.} \label{Table7}
\begin{tabular}{llllllllll}
\hline
 &  &    \multicolumn{2}{c}{\underline {\bf NGC 3211}} & \multicolumn{2}{c}{\underline{\bf NGC5979}}
    &    \multicolumn{2}{c}{\underline {\bf My60}} & \multicolumn{2}{c}{\underline {\bf M 4-2}}\\
Lab (\AA{})    & ID   & Model  & Observed & Model  & Observed & Model  & Observed & Model  & Observed \\
\hline
3425.88  & {[}Ne  V{]}   & 45.23 & 71.4$\pm$2.95 & 54.31 & 58.6$\pm$3.49 & 21.44 & 26.2$\pm$1.60 & 19.5  & 61.4$\pm$2.44  \\
3726.03 & {[}O   II{]}  & 33.32 & 17.4$\pm$0.66 & 18.2  & 3.4$\pm$0.19  & 15.10  & 5.2$\pm$0.36  & 15.34  & 11.8$\pm$0.43  \\
3728.82 & {[}O   II{]}  & 24.3 & 12.4$\pm$0.47 & 12.2  & 2.3$\pm$0.13  & 10.00  & 3.0$\pm$0.20  & 9.75  & 7.5$\pm$0.32  \\
3770.63 & HI   11-2     & 3.89  & 4.4$\pm$0.26  & 3.96  & 3.6$\pm$0.19  & 3.98  & 3.6$\pm$0.25  & 3.84  & 3.9$\pm$0.17  \\
3835.39 & HI   9-2      & 7.19  & 8.5$\pm$0.31  & 7.30  & 7.3$\pm$0.39  & 7.34  & 7.8$\pm$0.43  & 7.09  & 7.7$\pm$0.32  \\
3868.76 & {[}Ne  III{]} & 173.7 & 112.4$\pm$4.1 & 62.44 & 58.4$\pm$3.12 & 109.6 & 90.3$\pm$4.96 & 86.30 & 65.3$\pm$2.34  \\
3889.06 & HI   8-2      & 10.34 & 16.4$\pm$0.60 & 10.49  & 11.5$\pm$0.62& 10.55 & 15.8$\pm$0.85 & 10.2 & 14.5 $\pm$0.52  \\
3967.47 & {[}Ne  III{]} & 52.35 & 36.5$\pm$1.30 &  16.8 & 17.5$\pm$0.91& 33.02 & 22.9$\pm$1.24 &  26.01 & 15.9 $\pm$0.56  \\
3970.08 & HI   7-2      & 15.68 & 15.8$\pm$0.56 & 15.89 & 14.3$\pm$0.75& 15.99 & 15.3$\pm$0.80 & 15.46 & 15.5 $\pm$0.54  \\
4068.60 & {[}S   II{]}   & 1.01  & 1.8$\pm$0.10  & 0.20 & 1.3$\pm$0.07  &  0.20  & 2.0$\pm$0.16  & 0.34  & 2.4$\pm$0.10  \\
4076.35 & {[}S   II{]}  & 0.32  & 0.4$\pm$0.04  & 0.07  & 0.1$\pm$0.02  &   0.06    &      ...         &        0.11    &       ...        \\
4267.14 & C    II       & 0.15  & 0.5$\pm$0.05  & 0.80  & 0.8$\pm$0.05  & 0.55  & 0.6$\pm$0.12  & 0.90  & 0.9$\pm$0.08  \\
4340.47 & HI   5-2      & 46.98 & 49.2$\pm$1.52 &  47.11 & 44.4$\pm$2.05& 47.16 & 48.3$\pm$2.30 & 47.05 & 47.2 $\pm$1.47  \\
4363.21 & {[}O   III{]} & 22.75 & 25.1$\pm$0.81 & 14.78 & 14.7$\pm$0.70& 18.01 & 18.9$\pm$0.92 & 14.91 & 17.6 $\pm$0.56  \\
4471.50  & He   I        & 1.98  & 1.8$\pm$0.10  &0.70  & 0.8$\pm$0.05  & 2.11  & 2.2$\pm$0.17  & 1.32  & 1.3$\pm$0.12  \\
4541.59 & He   II       & 2.65  & 2.9$\pm$0.16  & 3.59  & 3.7$\pm$0.17  & 2.68  & 2.9$\pm$0.18  & 0.88  & 2.9$\pm$0.11  \\
4685.70 & He   II       & 86.21 & 85.3$\pm$2.52 & 105.1 & 105.5$\pm$4.67& 78.48 & 80.4$\pm$3.59 & 84.93 & 83.8$\pm$2.47  \\
4711.26 & {[}Ar  IV{]}  & 8.20 & 11.0$\pm$0.33 & 8.56 & 14.2$\pm$0.63&7.17  & 10.7$\pm$0.47 & 5.46  & 7.5$\pm$0.28  \\
4714.50 & {[}Ne  IV{]}  & 0.68  & 1.5$\pm$0.08  & 0.32  & 1.2$\pm$0.05  &   0.34       &       ...        & 1.86  & 1.0$\pm$0.09  \\
4724.89 & {[}Ne  IV{]} & 1.24  & 2.0$\pm$0.11  & 0.52  & 1.9$\pm$0.08  & 0.57  & 1.3$\pm$0.10  & 1.20  & 1.4$\pm$0.13  \\
4740.12 & {[}Ar  IV{]}  & 6.90  & 9.6$\pm$0.28  & 7.45 & 12.1$\pm$0.53 & 6.46  & 9.4$\pm$0.41  & 6.90  & 6.5$\pm$0.24  \\
4861.33 & HI   4-2      & 100.0 & 100.0$\pm$0.0 & 100.0 & 100.0$\pm$0.0 & 100.0 & 100.0$\pm$0.0 & 100.0 & 100.0$\pm$0.0  \\
4958.91 & {[}O   III{]} &  399.1 &528.5$\pm$14.6 & 312.0  &304.0$\pm$12.7 & 396.0 &422.6$\pm$17.7& 237.7 & 304.0$\pm$8.4  \\
5006.84 & {[}O   III{]} &1153.9 &1512$\pm$42   &901.8   &905.8$\pm$37   & 1145.0  & 1184.9$\pm$47 & 687.0   & 870.8$\pm$24 \\
5411.52 & He   II       & 6.07  & 6.3$\pm$0.17  & 8.26  & 8.6$\pm$0.33  & 6.17  & 6.3$\pm$0.24  & 2.02  & 6.3$\pm$0.21  \\
5517.71 & {[}Cl  III{]} & 1.0  & 1.0$\pm$0.10  & 0.50  & 0.5$\pm$0.03  & 0.49  & 0.5$\pm$0.05  & 0.43  & 0.4$\pm$0.04  \\
5537.87  & {[}Cl  III{]} & 0.90  & 0.9$\pm$0.09  & 0.48  & 0.5$\pm$0.03  & 0.51  & 0.5$\pm$0.05  & 0.42  & 0.4$\pm$0.04  \\
5875.66 & He   I        & 5.74  & 4.8$\pm$0.13  & 2.02 & 2.0$\pm$0.06  & 6.14  & 6.2$\pm$0.18  & 3.86  & 3.9$\pm$0.10  \\
6086.97  & {[}Fe  VII{]} & 0.11  & 0.1$\pm$0.02  & 0.09  & 0.1$\pm$0.00  & 0.10  & 0.1$\pm$0.01  & 0.10  & 0.1$\pm$0.02  \\
6101.83 & {[}K   IV{]}  & 0.53  & 0.5$\pm$0.05  & 0.70  & 0.7$\pm$0.03  & 0.6  & 0.5$\pm$0.02  & 0.50  & 0.5$\pm$0.02  \\
6300.30  & {[}O   I{]}   & 0.20  & 0.2$\pm$0.02  &   0.04     &    ...           &   0.02    &     ...          & 0.73  & 0.8$\pm$0.02  \\
6312.06  & {[}S   III{]} & 8.30 & 3.2$\pm$0.09  & 1.82 & 1.8$\pm$0.05  & 2.21  & 1.6$\pm$0.05  & 2.75  & 1.8$\pm$0.03  \\
6363.78 & {[}O   I{]}   & 0.08  & 0.1$\pm$0.02  &      0.01         &     ...          &      0.01         &    ...           & 0.24  & 0.3$\pm$0.03  \\
6433.12 & {[}Ar  V{]}   & 2.11  & 1.0$\pm$0.10  & 3.79  & 1.7$\pm$0.05  & 2.07  & 0.8$\pm$0.03  & 0.31  & 1.0$\pm$0.03  \\
6548.05  & {[}N   II{]}  & 4.22  & 5.8$\pm$0.17  & 0.78  & 0.7$\pm$0.03  & 1.22 & 1.1$\pm$0.03  & 3.72  & 4.2$\pm$0.07  \\
6560.09  & He   II       & 10.38 & 9.7$\pm$0.28  & 14.2 & 12.7$\pm$0.39 & 10.61 & 9.5$\pm$0.27  & 10.2 & 11.5$\pm$0.19  \\
6562.82 & HI   3-2      & 288.3 &286.7$\pm$3.4   & 284.0 & 274.7$\pm$4.3 & 282.1  &283.1$\pm$3.1  & 286.7 & 278.8$\pm$2.6  \\
6583.45 & {[}N   II{]}  & 12.43 & 15.5$\pm$0.45 & 2.31  & 2.4$\pm$0.07  & 3.58 & 3.5$\pm$0.10  & 10.92 & 10.4$\pm$0.17  \\
6678.15 & He   I        & 0.93  & 1.3$\pm$0.13  & 0.49 & 0.5$\pm$0.03  &1.47  & 1.6$\pm$0.05  & 0.90  & 1.0$\pm$0.02  \\
6683.45  & He   II       & 0.53  & 0.5$\pm$0.05  & 0.72 & 0.8$\pm$0.04  & 0.54  & 0.6$\pm$0.02  & 0.5  & 0.6$\pm$0.03  \\
6716.44 & {[}S   II{]}  & 3.90  & 2.7$\pm$0.15  & 0.68  & 0.6$\pm$0.03  & 0.59  & 0.6$\pm$0.02  & 0.89  & 1.3$\pm$0.02  \\
6730.82 & {[}S   II{]}  & 4.86  & 3.2$\pm$0.10  & 0.93  & 0.7$\pm$0.03  & 0.86  & 0.8$\pm$0.03  &1.19 & 1.7$\pm$0.03  \\
6795.16  & {[}K   IV{]}  & 0.1  & 0.1$\pm$0.01  & 0.21  & 0.2$\pm$0.01  & 0.1  & 0.1$\pm$0.01  & 0.03  & 0.1$\pm$0.02  \\
6890.91 & He   II       & 0.68  & 0.7$\pm$0.07  &0.92  & 1.0$\pm$0.05  & 0.69  & 0.7$\pm$0.02  & 0.22  & 0.8$\pm$0.03  \\
7005.83 & {[}Ar  V{]}   & 4.61 & 1.8$\pm$0.10  & 8.05  & 3.5$\pm$0.11  & 4.41 & 1.8$\pm$0.05  & 11.97   & 2.1$\pm$0.05 \\

\hline
\end{tabular}
\end{table*}

\subsection{Distances from Photoionisation Models}
In this section, we will develop on a method to estimate distances
based only upon the requirement that the model reproduce both the
observed flux and the observed angular size of the PN. The success
of this method depends critically on how well we have determined
$f_{\rm S}$, since this will in turn determine the relationship
between the luminosity of the central star, and the absolute
H$\beta$ luminosity of the nebula. Our model, in essence, consists
of the application of simple Str\"omgren theory. For a given
electron density $n_e$ and nebular radius, $r$, the H$\beta$
luminosity $L_{\rm H\beta}$ is given by:
\begin{equation}
L_{\rm H\beta} \propto n_e^2 r^3 \label{Lb}
\end{equation}
and the angular radius $\theta$ is given in terms of the distance $d$, $\theta =r/d$. The reddening-corrected observed H$\beta$ luminosity $F_{\rm H\beta}$ is given by:
\begin{equation}
F_{\rm H\beta} = L_{\rm H\beta} /(4\pi d^2) \label{Fb}
\end{equation}
Our photoionisation models give (for any assumed distance and
stellar luminosity $L_*$) the values of $L_{\rm H\beta}, r$ and
$n_e$, and the observations give us $F_{\rm H\beta}$ and $\theta$,
so from these equations we can solve for the distance at which the
model predicted angular radius $\theta$ and H$\beta$ flux agree with
the observed values. Note that the solution depends critically on
the accurate determination of the parameter  $f_{\rm S}$, which
affects both the absolute flux and the radius of the model, and on
$\log P/k$, which also has a strong effect on the radius predicted
by the model for any assumed stellar luminosity. Observationally,
the error in the determination of  $F_{\rm H\beta}$ is not of
concern, but the error in the determination of $\theta$ has a much
greater effect on the solution, since the nebular boundary is not
always clearly defined. Where possible, we have used the angular
radius for the optically-thick part of the nebula, as measured in
the [N\,II] line, but for NGC\,5979 and My\,60 we have used the
angular diameter as measured in the H$\alpha$ line.

\begin{figure}
\includegraphics[scale=0.4]{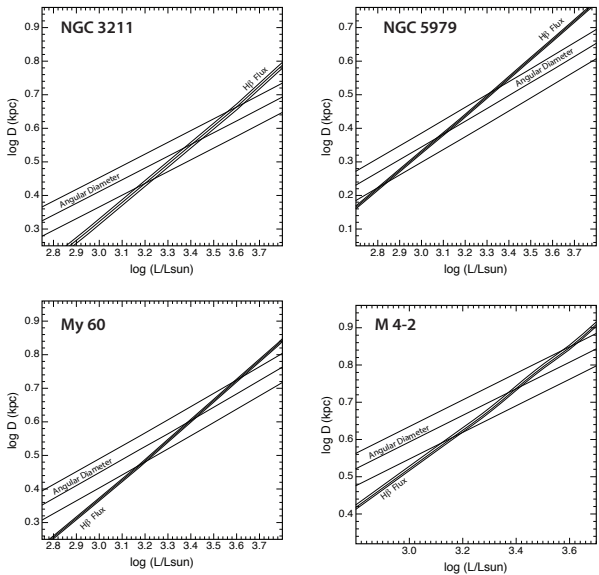}
 \caption{The solution to the absolute luminosity and distance of the PNe given by the requirement that both
 the reddening-corrected observed H$\beta$ flux and the observed angular size of the PNe agree with those
 predicted by the photoionisation model. The error bands are the measurement errors of the observed quantities.}
  \label{fig7}
\end{figure}

The solutions for the distance of each PN are given graphically in
Figure \ref{fig7}. The distances derived by this method are as
follows, where the mean of the statistical distance scale estimates
is given in parentheses. NGC\,3211 $3.6^{+1.05}_{-0.96}$\,kpc
(2.61\,kpc);  NGC\,5979  $2.4^{+0.9}_{-0.6}$\,kpc (2.99\,kpc);
My\,60 $4.0^{+1.1}_{-0.8}$\,kpc (3.76\,kpc);  M\,4-2
$5.5^{+1.7}_{-1.5}$\,kpc (5.90\,kpc). In each case the agreement
with the statistical distance estimate is within the error bar. This
gives confidence in the validity of the method. Nonetheless, due to
both the modelling errors and the observational limitations
mentioned above, we conclude that this method cannot deliver
sufficient accuracy to supplant the statistical distance estimate
techniques, but is certainly sufficient to provide an independent
check on these.

At these distances, the nebular diameter (D) as given from the
photoionisation model are a follows; NGC\,3211 $D= 0.138$pc,
NGC\,5979 $D=0.174$pc, My\,60 $D=0.153$pc, and  M\,4-2 $D=0.237$pc.
In the case of NGC\,3211 the diameter is measured to the edge of the
[N\,II] zone, while in the other three objects the diameter refers
to the diameter of the optically-thin region.

\subsection{Location of CSs on the H-R Diagram}
With a reasonably accurate knowledge of the distance, we can now
place the central stars on the Hertzsprung-Russell (H-R) Diagram.
For this purpose, we adopt a mean of the statistical and
photoinisation distances: NGC\,3211 $3.2^{+1.2}_{-1.0}$\,kpc;
NGC\,5979  $2.7^{+1.0}_{-0.7}$\,kpc; My\,60
$3.9^{+1.1}_{-0.8}$\,kpc;  M\,4-2  $5.8^{+1.7}_{-1.5}$\,kpc. With
these distances, we derive stellar luminosities and their errors,
and from the models, we also provide the derived stellar
temperatures with their estimated measurement errors. These are
given in Table \ref{Table8}.

\begin{table}
\centering
\caption{Derived Central star parameters.} \label{Table8}
\begin{tabular}{|l|c|c|}
  \hline
   & log L ($L_{\odot}$)  & log $T_{\rm eff}$ K   \\
\hline \\
 NGC 3211 & $3.30^{+0.13}_{-0.13}$ & $5.16^{+0.03}_{-0.03}$ \\
 NGC 5979 & $3.18^{+0.15}_{-0.15}$ & $5.20^{+0.03}_{-0.03}$ \\
  My 60 & $3.39^{+0.17}_{-0.17}$ & $5.11^{+0.03}_{-0.03}$ \\
  M 4-2 & $3.44^{+0.20}_{-0.18}$ & $5.13^{+0.03}_{-0.03}$ \\
  \hline
\end{tabular}
\end{table}

\begin{figure}
\includegraphics[scale=0.65]{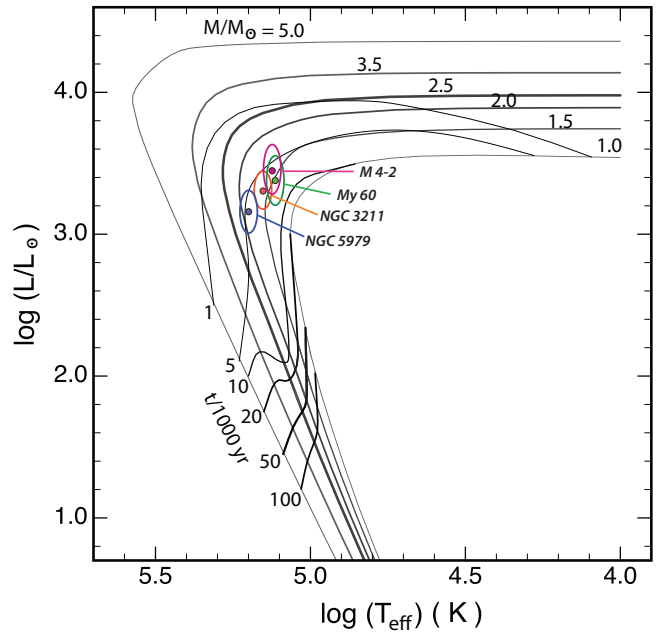}
 \caption{The location of our PNe on the (hydrogen-burning) H-R diagram as derived from photoionisation modelling
 compared with evolutionary tracks of central stars of PNe from \citet{VW94}. The four objects are remarkably similar,
 and consistent with having initial masses in the range $1.5-2.0 {\rm M}_{\odot}$, and evolutionary ages of about
 6000yr.}
\label{fig8}
\end{figure}

In Figure \ref{fig8} we locate these stars on the Hertzprung-Russell
diagram. Overlaid are  the H-burning $Z=0.016$ tracks from
\citet{VW94}. What is remarkable is the manner in which the four
objects of our study cluster tightly together, suggesting very
similar precursor stars. All are consistent with having initial
masses between 1.5 and 2.0\,M$_{\odot}$, and having an evolutionary
age of about 6000\,yr. It is interesting that the only bona-fide
double-shell PN, NGC\,5979, seems to have the highest mass
precursor. This would be consistent with the evolutionary scenario
of \citet{VW93}, in that the He-shell flashes occur more frequently
in higher mass precursors, driving ``super-wind'' episodes which
then form multiple shells in the subsequent PNe.

The age of the PNe as measured from the H-burning tracks is in fact
the time since the central star became hotter than $10^4$K. This can
be directly compared to the estimated dynamical ages given from the
diameters given above, and the expansion velocity in the
low-ionisation species as given in Table \ref{Table6}. These are
respectively; NGC\,3211 $\tau_{\rm dyn}= 3000$yr, NGC\,5979
$\tau_{\rm dyn}= 3950$yr, My\,60 $\tau_{\rm dyn}= 3100$yr, and
M\,4-2 $\tau_{\rm dyn}= 4500$yr. The dynamical ages are somewhat
shorter than the age implied from the evolutionary tracks, but both
methods agree that the ages of these four PNe are very similar. Due
to acceleration or deceleration of the nebular shell, driven by the
changing stellar wind parameters during its dynamical evolution, we
have no reason to expect that the dynamical age derived will be
exactly equal to the age inferred by the position of the PN on the
evolutionary track. Thus we see no serious discrepancy between the
dynamical and the evolutionary age. If we were to use the recent
models of \citet{Bertolami16}, then the observed dynamical ages
would show closer agreement to the theoretical ages derived from the
position of PNe along the post-AGB track.

\section{Conclusions}
In this paper we have studied four PNe, three of which had been
classified with central  stars of the WELS type, and the fourth
which showed the same set of recombination lines of highly-excited
ions which are thought to characterise a WELS;  C IV at 5801-12\AA{}
and the N III, C III, and C IV complex feature at 4650\AA{}. We
have, however, demonstrated that these emission lines arise not in
the central star, but are instead distributed throughout the nebular
gas. We find no trace of these emission originating from the central
stars themselves. Instead, we identify O\,VI $\lambda\lambda 3811.3,
3834.2$ in three of the central stars, and in the case of NGC\,5979
the  O\,V $\lambda 4664.4$ line as well. Thus, at least three of the
central stars are not WELS, but are in fact hot O(H) stars. This
result casts doubt on the reality of the WELS class as a whole, a
result made possible only by the fact that we have integral field
spectroscopy covering the whole nebula.

We cannot use nebular emission line characteristics to define a
class of the central star, since the nebular lines depend on the
nature of the EUV spectrum, the chemical abundances, the density and
the geometry of the gas with respect to the central star. We always
define the central star type itself from the UV, visible or IR
emissions of the central star. If the emission lines which have been
used to classify the star instead arise in the nebula, as we have
demonstrated here in four independent examples, then the
classification has no validity. Clearly, as in the case of binary
systems, some real examples of WELS exist. However, given that we
have found four out of four examples of misclassification, we
believe that many other cases of such misclassification may exist.
This is the major conclusion of this paper.

As to why the C IV at 5801-12 \AA{}  and the N III, C III, and C IV
complex  feature at 4650 \AA{} are so strong, we cannot provide a
solution, since we do not have the required atomic data relating to the
effective recombination coefficients. It may be possible that we have
C and N- rich pockets of gas in the nebula, but then why would these
species be distributed in exactly the way we would expect in terms
of the local excitation? Figures \ref{fig1} and \ref{fig3} are
remarkably similar.

Apart from the standard nebular analysis techniques used to probe
chemical  abundances and physical conditions in the nebula, we have
been able to apply self-consistent photoionisation modelling to an
analysis of the integrated spectrum of these PNe. We find that an
optically-thin isobaric model with an inner empty zone and, in some
cases, an additional optically-thick zone provides an excellent
description of the observed spectrum, and provides a good estimate
of the stellar effective temperature. Generally speaking, the
agreement between the two analysis methods is quite satisfactory as
far as the abundance determinations are concerned.

According to the derived chemical abundances from the NEAT, none of
our objects was classified as Type I PN. We find that the chemical
compositions of both NGC 3211 and NGC 5979 are consistent with being
of Type IIa, while that of My 60 and M 4-2 are consistent with the
Type IIb/III classification. Considering also their calculated
Galactic vertical heights and peculiar velocities, we suggest that
NGC 3211 is probably a member of the Galactic thin disk, while both
My 60 and M 4-2 are Galactic thick members.

Additionally, the self consistent modelling yields a number of
additional parameters.  Specifically, we are able to use the models
to provide a new ``Str\"omgren Distance'' estimate, which agrees -
within the errors - with the distances derived from the standard
statistical techniques. We find that all four PNe are highly-excited
nebulae optically-thin to the escape of EUV photons. With a
knowledge of the distance, another parameters derived from the
models, we are also able to constrain the luminosity of the central
star, placing it on the H-R Diagram.

All four PNe studied in this paper are of remarkably similar $T_{\rm
eff}$ and luminosity,  are at a very similar evolutionary stage with
ages in the range 3000-6000\,yr, and are all derived from precursor
stars in the restricted mass range $1.5-2.0$\,M$_{\odot}$.

\section*{acknowledgements}
The authors would like to thank the anonymous referee for valuable
and constructive comments.

\bibliographystyle{mn2e_new}
\bibliography{MN2016}

\appendix
\section{Measured Emission Line Fluxes}
We determined the final integrated emission-line fluxes from the
combined,  flux-calibrated blue and red spectra of each PN. The line
fluxes and their measurement uncertainties were determined using the
IRAF {\tt splot} \footnote{IRAF is distributed by the National
Optical Astronomy Observatory, which is operated by the Association
of Universities for Research in Astronomy (AURA) under a cooperative
agreement with the National Science Foundation.} task and were
integrated between two given limits, over a local continuum fitted
by eye, using multiple Gaussian fitting for the line profile.

The amount of interstellar reddening was determined by the NEAT code
from  the ratios of hydrogen Balmer lines, in an iterative method.
Firstly, the reddening coefficient, $c(H\beta)$, was calculated
assuming intrinsic $H\alpha$, $H\beta$, $H\gamma$ line ratios for a
temperature ($T_e$) of 10000K  and density ($N_e$) of 1000
cm$^{-3}$, which were used to de-reddened using this value the line
list. Secondly, the electron temperature and density were calculated
as explained in section (3.2). After that, the intrinsic Balmer line
ratios were re-calculated at the convenient temperature and density,
and again $c(H\beta)$ was recalculated. The line intensities have
been corrected for extinction by adopting the extinction law of
Howarth (1983). The estimated reddening coefficients and observed
$H\beta$ fluxes of the sample are listed in Table \ref{Table2},
while Table \ref{TableA1} lists the observed and de-reddened line
strengths of the sample. Columns (1) \& (2) give the laboratory
wavelengths and identification of observed emission lines, while
columns (3)-(10) give the observed F($\lambda$) and the de-reddened
line strengths I($\lambda$), scaled to $H\beta =100$.  A total of
$\sim$ 128 distinct emission lines were identified in each nebula,
which include $\sim$ 88 permitted and 40 forbidden lines.

\onecolumn
\begin{table}
\caption{Line fluxes and de-reddened intensities, relative to
H$\beta = 100$, of NGC 3211, NGC 5979, My 60, and M 4-2. The full
version of the table is available online. A portion is shown here
for guiding the reader regarding its content.}
\begin{tabular}{llcccccccc}
\hline
 &  &    \multicolumn{2}{c}{\underline {\bf NGC 3211}} & \multicolumn{2}{c}{\underline{\bf NGC 5979}}
    &    \multicolumn{2}{c}{\underline {\bf My 60}} & \multicolumn{2}{c}{\underline {\bf M 4-2}}\\
Lab (\AA{})    & ID   & F($\lambda$)  & I($\lambda$) & F($\lambda$) & I($\lambda$) & F($\lambda$) & I($\lambda$) & F($\lambda$) & I($\lambda$)\\
\hline \\
3868.76 & {[}Ne  III{]} & 92.5$\pm$1.85 & 112.4$\pm$4.1 & 47.1$\pm$1.41 & 58.4$\pm$3.12 & 57.2$\pm$1.72 & 90.3$\pm$4.96 & 48.1$\pm$0.96 & 65.3$\pm$2.34  \\
3889.06 & HI   8-2      & 13.6$\pm$0.27 & 16.4$\pm$0.60 & 9.3$\pm$0.28  & 11.5 $\pm$0.62& 10.1$\pm$0.30 & 15.8$\pm$0.85 & 10.7$\pm$0.21 & 14.5 $\pm$0.52  \\3923.48 & He   II       & 0.4$\pm$0.04  & 0.5$\pm$0.05  & 0.6$\pm$0.02  & 0.8$\pm$0.04  &               &               & 0.5$\pm$0.01  & 0.7$\pm$0.03  \\
3967.47 & {[}Ne  III{]} & 30.6$\pm$0.61 & 36.5$\pm$1.30 & 14.4 $\pm$0.43& 17.5 $\pm$0.91& 15.1$\pm$0.45 & 22.9$\pm$1.24 &  12.0$\pm$0.24 & 15.9 $\pm$0.56  \\
3970.08 & HI   7-2      & 13.2$\pm$0.26 & 15.8$\pm$0.56 & 11.8 $\pm$0.35& 14.3 $\pm$0.75& 10.1$\pm$0.30 & 15.3$\pm$0.80 & 11.8$\pm$0.24 & 15.5 $\pm$0.54  \\
4025.61  & He   II       &               &               &               &               &               &               & 1.5$\pm$0.05  & 2.0$\pm$0.08  \\
4026.08 & N    II       & 1.5$\pm$0.15  & 1.8$\pm$0.19  & 1.3$\pm$0.04  & 1.5$\pm$0.08  & 1.3$\pm$0.13  & 1.9$\pm$0.21  &               &              \\
4068.60 & {[}S   II{]}   & 1.5$\pm$0.08  & 1.8$\pm$0.10  & 1.1$\pm$0.03  & 1.3$\pm$0.07  &  1.4$\pm$0.10  & 2.0$\pm$0.16  & 1.9$\pm$0.06  & 2.4$\pm$0.10  \\
\hline
\end{tabular}
\label{TableA1}\\
\end{table}
\end{document}